\documentclass[11pt]{article}

\usepackage[ruled,vlined,linesnumbered,noresetcount]{algorithm2e}
\makeatletter
\newcommand{\AlgoResetCount}{\renewcommand{\@ResetCounterIfNeeded}{\setcounter{AlgoLine}{0}}}
\newcommand{\AlgoNoResetCount}{\renewcommand{\@ResetCounterIfNeeded}{}}
\newcounter{AlgoSavedLineCount}

\newcommand{\QED}{\hfill\ensuremath{\square}}
\makeatother

\usepackage{graphicx, graphics}
\graphicspath{{./plots/}}

\usepackage{enumerate}
\usepackage{courier}
\usepackage{amssymb,amsmath}

\usepackage{color}
\definecolor{dkgreen}{rgb}{0,0.6,0}
\definecolor{grey}{rgb}{0.4,0.4,0.4}

\usepackage{hyperref}

\usepackage[hang, small]{caption}
\usepackage[caption=false]{subfig}
\usepackage{titlesec}
\titleformat*{\section}{\Large \bf}
\titleformat*{\subsection}{\large \bf}
\usepackage{array}
\usepackage{booktabs, tabularx} 
\newcolumntype{Y}{>{\centering\arraybackslash}X}

\newcommand{\mt}{\ensuremath{\mathbf{t}}}
\newcommand{\mtheta}{\ensuremath{\boldsymbol{\theta}}}
\newcommand{\mDelta}{\ensuremath{\boldsymbol{\Delta}}}
\newcommand{\mM}{\ensuremath{\mathbf{M}}}
\newcommand{\mv}{\ensuremath{\mathbf{v}}}
\newcommand{\ga}{\ensuremath{G_{+}}}
\newcommand{\gd}{\ensuremath{G_{-}}}

\definecolor{olive}{RGB}{0,151,0}

\usepackage{authblk}

   \makeatletter
\def\@fnsymbol#1{\ensuremath{\ifcase#1\or *\or
   \mathsection\or \mathparagraph\or \|\or **\or \dagger\dagger
   \or \ddagger\ddagger \else\@ctrerr\fi}}
    \makeatother

\begin{document}

\title{Local Graph Stability in Exponential Family Random Graph Models\thanks{This work was supported by NSF awards DMS-1361425 and SES-1826589.}}
\author[1]{Yue Yu\thanks{yuey6@uci.edu}}
\author[2,3]{Gianmarc Grazioli\thanks{g.grazioli@uci.edu}}
\author[4]{Nolan E. Phillips\thanks{nolanephillips@fas.harvard.edu}}
\author[1,2,5]{Carter T. Butts\thanks{buttsc@uci.edu}\thanks{To whom correspondence should be addressed: \texttt{buttsc@uci.edu}}}
\affil[1]{Department of Computer Science}
\affil[2]{California Institute for Telecommunications and Information Technology}
\affil[3]{Department of Chemistry}
\affil[4]{Department of Sociology, Harvard University}
\affil[5]{Departments of Sociology, Statistics, and Electrical Engineering and Computer Science; and Institute for Mathematical Behavioral Sciences, University of California, Irvine}
\date{}
\maketitle

\abstract{Exponential family Random Graph Models (ERGMs) are widely used to model networks by parameterizing graph probability in terms of a set of user-selected sufficient statistics. Equivalently, ERGMs can be viewed as expressing a probability distribution on graphs arising from the action of competing social forces that make ties more or less likely, depending on the state of the rest of the graph.  Such forces often lead to a complex pattern of dependence among edges, with non-trivial large-scale structures emerging from relatively simple local mechanisms.  While this provides a powerful tool for probing macro-micro connections, much remains to be understood about how local forces shape global outcomes.  One very simple question of this type is that of the conditions needed for social forces to stabilize a particular structure: that is, given a specific structure and a set of alternatives (e.g., arising from small perturbations), under what conditions will said structure remains more probable than the alternatives?  We refer to this property as local stability and seek a general means of identifying the set of parameters under which a target graph is locally stable with respect to a set of alternatives.  Here, we provide a complete characterization of the region of the parameter space inducing local stability, showing it to be the interior of a convex cone whose faces can be derived from the change-scores of the sufficient statistics vis-\`a-vis the alternative structures.  As we show, local stability is a necessary but not sufficient condition for more general notions of stability, the latter of which can be explored more efficiently by using the ``stable cone'' within the parameter space as a starting point.  In addition to facilitating the understanding of model behavior, we show how local stability can be used to determine whether a fitted model implies that an observed structure would be expected to arise primarily from the action of social forces, versus by merit of the model permitting a large number of high probability structures, of which the observed structure is one (i.e. entropic effects). We also use our approach to identify the dyads within a given structure that are the least stable, and hence predicted to have the highest probability of changing over time.}

\section*{Notation} 
We denote vectors by lower case boldface letters, as in $\mt$, matrices by uppercase boldface letter, as in \mM. $|S|$ is the cardinality of a set $S$.


\section{Introduction}

The last several decades have seen a growth of interest in network data and in strategies to analyze such data. 
While statistical analysis of social networks goes back to the 1930s \cite{moreno.jennings:soc:1938}, advances in computing and statistical theory have fueled a particular rise in stochastic models for networks with complex structure  \cite{watts2004new}. The importance of capturing nontrivial aspects of network structure has been motivated by studies of phenomena such as information transmission (e.g. \cite{boorman1975combinatorial,dodds2003information,cowan2004network}), systemic robustness (e.g. \cite{krackhardt.stern:spq:1988,callaway2000network,klau.weiskircher:ch:2005,acemoglu2015systemic}), and disease transmission (e.g. \cite{morris1997concurrent, Morris2009concurrent}), all of which can be significantly impacted by features such as clustering and community or subgroup structure that are not readily reproduced by simple random graph models.

Another motivation for complex network models has been the elucidation of the connection between local and global aspects of network structure.  For instance, the frequency distribution of triadic subgraphs strongly constrains higher-order structures like ranked clusters \cite{holland.leinhardt:cgs:1971}, and partnership concurrency is closely related to forward connectivity in time-varying networks \cite{Morris2009concurrent}.  Biases in subgraph frequencies are themselves directly related to the conditions under which the state of one edge depends upon another \cite{frank.strauss:jasa:1986,pattison.robins:sm:2002,snijders.et.al:sm:2006}, creating a direct link between local processes that e.g. favor or inhibit tie formation or triadic closure and higher-order structure.  The exponential family random graph modeling (ERGM) framework (discussed in detail below) has become a widely used approach for identifying and exploring such connections between local and global structure in social and other networks \cite{robins.et.al:ajs:2005,lusher.et.al:bk:2012}.  While it can be seen simply as a flexible language for specifying distributions on graph sets, ERGMs can also be interpreted as parameterizing a set of biases influencing relational structure, with realized networks emerging from the interplay of these biases; these biases are formally analogous to forces in a physical context, an analogy that has been exploited in applications of ERGMs to biophysical systems (e.g. \cite{grazioli.et.al:jpcB:2019,grazioli.et.al:fmb:2019}).  In some cases, these biases can also be interpreted in terms of utility theory \cite{snijders:sm:2001}, with network structure arising from the equilibrium of a latent stochastic choice process in which agents' decisions to add or remove ties are shaped by the associated biases.  One potential use of ERGMs is hence to probe the conditions that are sufficient for the emergence or persistence of particular types of network structure, particularly where multiple mechanisms may be simultaneously at work.

In this paper, we examine one facet of this latter question, specifically introducing a basic notion of local \emph{network stability} vis a vis an ERGM family and characterizing the subspace of parameters for an arbitrary family that renders a target structure stable with respect to a set of alternative networks.  As we show, the stabilizing region of the parameter space forms the interior of a convex cone originating at the origin, whose faces are associated with a subset of the alternative networks against which the target is being compared.  These results are presented in section \ref{sec:stability}, along with a practical algorithm for efficiently finding the stabilizing region of the parameter space. Our stability analysis is then illustrated with a simple and intuitive example involving an idealized centralized group structure (section \ref{sec:starStructure})), followed by an application to a well-known study of collaboration within a legal firm (in section \ref{sec:lazega}).  In both cases, the correspondence between our notion of local stability and persistence of structures under Metropolis dynamics (widely employed in Markov Chain Monte Carlo simulations of network structure) is explored, and the use of stability calculations to predict likely or unlikely edge changes is demonstrated.



\section{Stability}
\label{sec:stability}

Here we introduce a simple notion of local stability for network structures. The intuition is as follows.  Assume we have a graph whose stability is to be assessed (the \emph{target network} or \emph{target graph}) with respect to a set of \emph{alternative graphs} that might be observed, as well as a model family whose elements are probability distributions on a graph set that includes the target graph and the alternative set.  Our goal is to find a subset of models within the model family under which the target graph is more probable than any of the graphs in the alternative set.  When this subset is non-empty, its members are said to \emph{stabilize} the target graph vis a vis the alternative set, and by turns the target graph is said to be \emph{locally stable} vis a vis the alternative set under any model within the subset.



It should be noted that our approach is very general: the target graph is given to us from the outset; there is no limitation on the choice of the alternative set; and although parts of our discussion require the model to have some specific functional properties, it is not limited to only the exponential family models.  However, we model families in ERGM form admit a particularly elegant characterization of stability, and we employ this framework throughout this paper.

In the remainder of this section, we first formally define our notion of stability, and show how the use of the ERGM formalism allows the stabilizing subset of models within a given family to be easily characterized.  We discuss some important properties of this stabilizing subset (including its geometric representation), and then provide an efficient approach to computing it in practice.

\subsection{Definitions}
\label{sec:def}
\begin{figure}[h!]
  \centering
  \includegraphics[width=10cm]{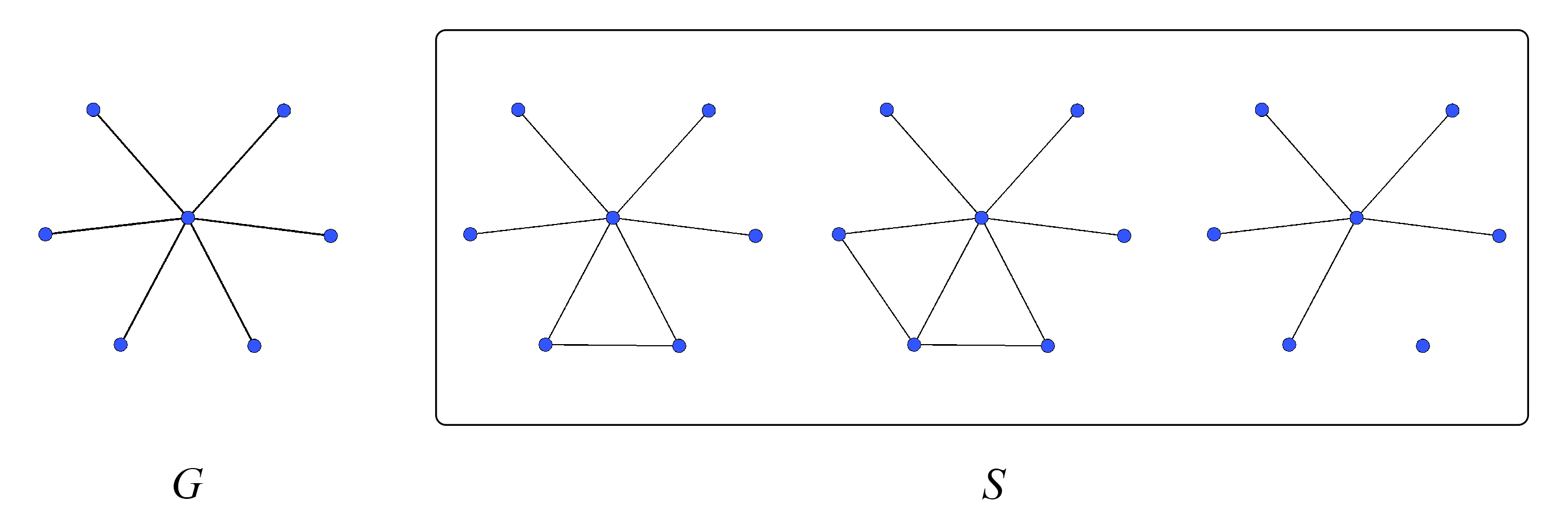}
  \caption{An example of a target graph $G$ and alternative set $S$.}
  \label{fig:stability}
\end{figure}
Let $\mathcal{G}$ be a finite set of graphs, $G\in\mathcal{G}$ be a \emph{target graph} whose stability is to be assessed, and $S\subseteq \mathcal{G}\setminus G$ an \emph{alternative set} of graphs with respect to which $G$ is intended to be stable. For clarity of illustration we will emphasize the case in which all graphs in $G \cup S$ share the same vertex set, although this is not assumed.  Likewise, we illustrate our ideas on simple graphs, but our development applies equally to directed, weighted, or multiplex networks.

We assess stability with respect to a model family on $\mathcal{G}$, which we take to be 
specified in ERGM form:
\begin{equation}
\Pr(G = g|\mtheta) = \frac{\exp{(\mtheta\cdot\mt(g))}}{\mathcal{K(\mtheta)}}, \text{ where } \mathcal{K}(\mtheta) = \sum_{g \in \mathcal{G}}e^{\mtheta^T\mt(g)}, 
\label{eq:ergm}
\end{equation}

where $\mtheta \in \mathbb{R}^K$ is a vector of parameters and $\mt:\mathcal{G}\mapsto \mathbb{R}^K$ is a vector of sufficient statistics.  Intuitively, each element of $\mt$ corresponds to a graph property whose distribution in $G$ is to be biased, with the corresponding element of $\mtheta$ indicating the strength and direction of the bias in question; in particular, $\mathbf{E}_{\mtheta} t_i(G)$ is monotone in $\mtheta_i$.  Our goal is then to find the set of coefficients $\Phi = \{\mtheta\}$ that make $G$ more probable than any of the graphs in $S$, i.e.
\begin{equation}
\mtheta \text{, such that } \Pr(G | \mtheta) > \Pr(G' | \mtheta), \text{ for } G' \in S
\label{eq:prob}
\end{equation}
with $\Phi=\{\varnothing\}$ if no $\mtheta$ satisfies this condition.  Any $\mtheta\in \Phi$ is said to \emph{stabilize} $G$ with respect to $S$, and $\Phi$ defines the \emph{stabilizing subfamily} of the model family parameterized by $\mt$ for $G$ with respect to $S$.  For convenience, we also refer to $\Phi$ as the \emph{stabilizing subspace} of the full parameter space, noting the equivalence of models and parameter vectors in this representation.

\subsubsection{Characterization of the Stable Subspace} \label{sec:propertisOfSR}

Using (\ref{eq:ergm}), we can rewrite the inequality in (\ref{eq:prob}) as
\begin{align*}
\frac{e^{\mtheta^T\mt(G)}}{\mathcal{K}(\mtheta)} > \frac{e^{\mtheta^T\mt(G')}}{\mathcal{K}(\mtheta)}
\end{align*}
allowing us to simplify the stability condition as
\begin{align}
\mtheta: \mtheta^T(\mt(G') - \mt(G)) < \mathbf{0} \text{ for } G' \in S. \label{eq:potent}
\end{align}

The quantity $\mtheta^T\mt(G)$ is called the ERGM \emph{potential}, and is equal to the log probability of $G$ up to an additive constant.  $\mDelta(G,G') = \mt(G') - \mt(G)$ is called the \emph{change score}, and describes the way in which the sufficient statistics differ between graphs.  
%
%
We may then construct a matrix $\mM$ by accumulating the change score vectors for $G$ versus all $G'$ in $S$.  Specifically, we define $\mM$ to be a $|S|\times K$ matrix where the $i$-th row is the change score $\Delta(G,G_i')$, and the $j$-th column is the change scores regarding the $j$-th sufficient statistics:
\begin{equation*}
M_{ij} =\Delta_{j}(G,G'_i), G'_i \in S.
\end{equation*}

With this notation it becomes clear that we can easily describe the local stability problem algebraically as
\begin{equation}
\begin{aligned}
&\text{Find all } \mtheta \\
&\text{Such that } \mM\mtheta = \mv \in \mathbb{R}^{|S|}_-
\end{aligned}
\label{eq:final}
\end{equation}
with the set of $\mtheta\in \mathbb{R}^K$ that satisfy the above constraint defined as $\Phi$.

\begin{figure}[h!]
  \centering
  \includegraphics[width=5cm]{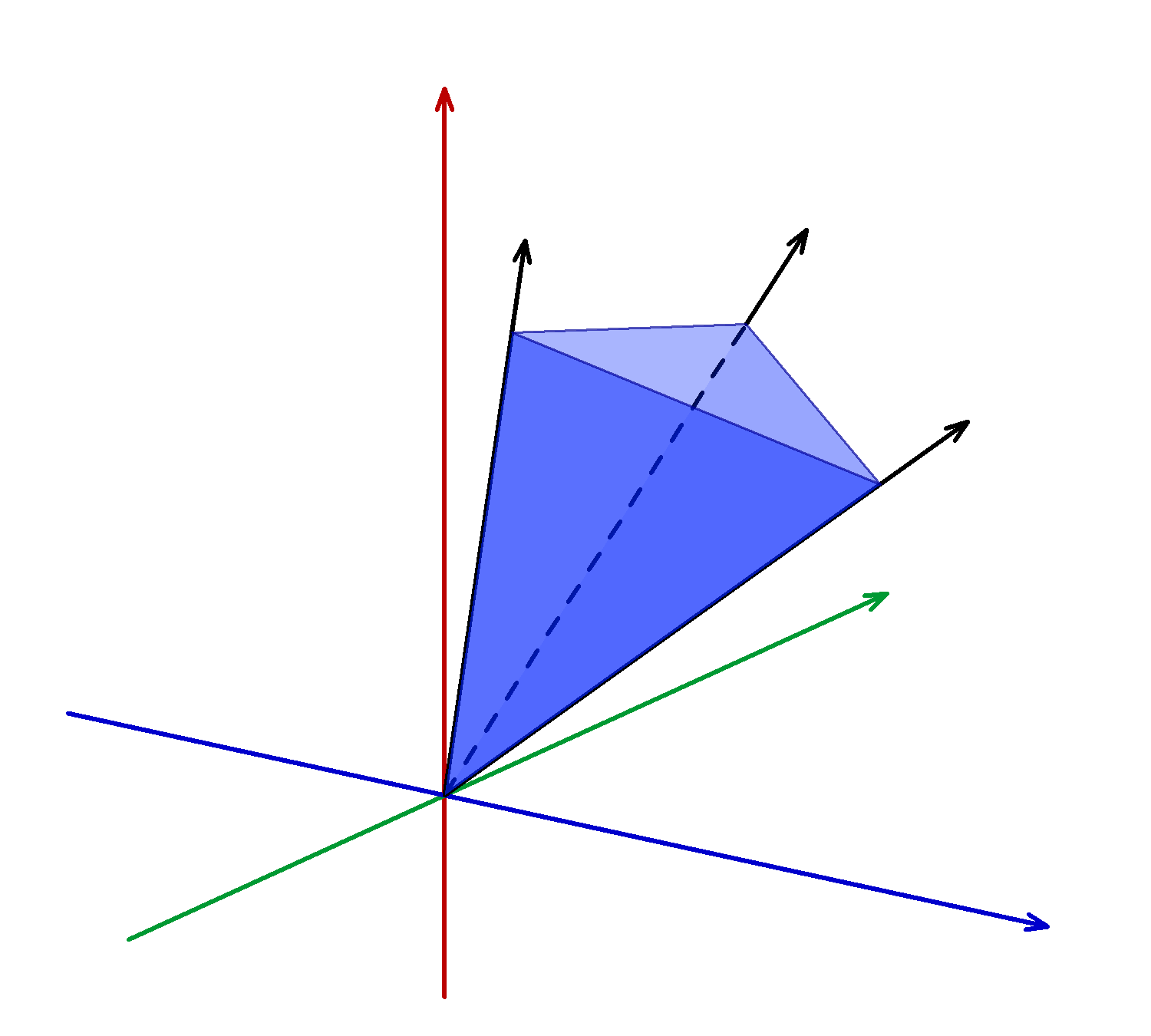}
  \caption{An example of the convex cone formed by 3 hyperplanes, in three-dimensional space.}
  \label{fig:convex_cone}
\end{figure}

The algebraic characterization of $\Phi$ immediately reveals several useful properties of the stabilizing subspace.
Trivially, the matrix product $\mM \mtheta$ in Eq.~\ref{eq:final} can be rewritten as a set of row-wise inner products arising from members of the comparison set:
\[\sum_j M_{ij} \theta_j < 0, \text{ for } 1<i<|S|.\] 
Each corresponding inequality represents an \emph{open half space}, whose \emph{dividing hyperplane} (eg. $\sum_j M_{ij}\theta_j = 0$) passes through the origin. Intuitively, each half-space represents the portion of the parameter space under which $G$ is more probable than a particular member of $S$.  The intersection of these open half-spaces, if nonempty, is the interior of a convex polytope cone emanating from the origin (Fig.~\ref{fig:convex_cone}). We call this convex cone the \emph{stable cone} of $G$ w.r.t the alternative set $S$, since any parameter vector within it stabilizes $G$.
 
We can also prove the set $\Phi$ is a convex cone by showing that if $\theta_1, \theta_2 \in \Phi$ then $\alpha\theta_1 + \beta\theta_2 \in \Phi$, for any $\alpha$ and $\beta > 0$.
\textit{Proof}: Let $v_1 = M\theta_1$ and $v_2 = M\theta_2$, then $v_1,v_2 \in \mathbb{R}^{|S|}_-$. So,
\begin{align*}
v' &= M[\alpha\theta_1 + \beta\theta_2] \\
  & = \alpha v_1 + \beta v_2 & \text{ because $\alpha, \beta >0$}&\\
& \in \mathbb{R}^{|S|}_- \\
 &\QED
\end{align*} 
A practical implication of this observation is that the stabilizing subspace can be characterized in terms of a set of vectors (the directions of the facial intersections of the stable cone), from which any member of $\Phi$ can be obtained.  In practice, we might expect that redundancies in the constraints implied by $M$ will limit the number of vectors that are needed, an idea that we exploit below.

\subsection{Local Stability}

\begin{figure}[h!]
  \centering
  \includegraphics[width=10cm]{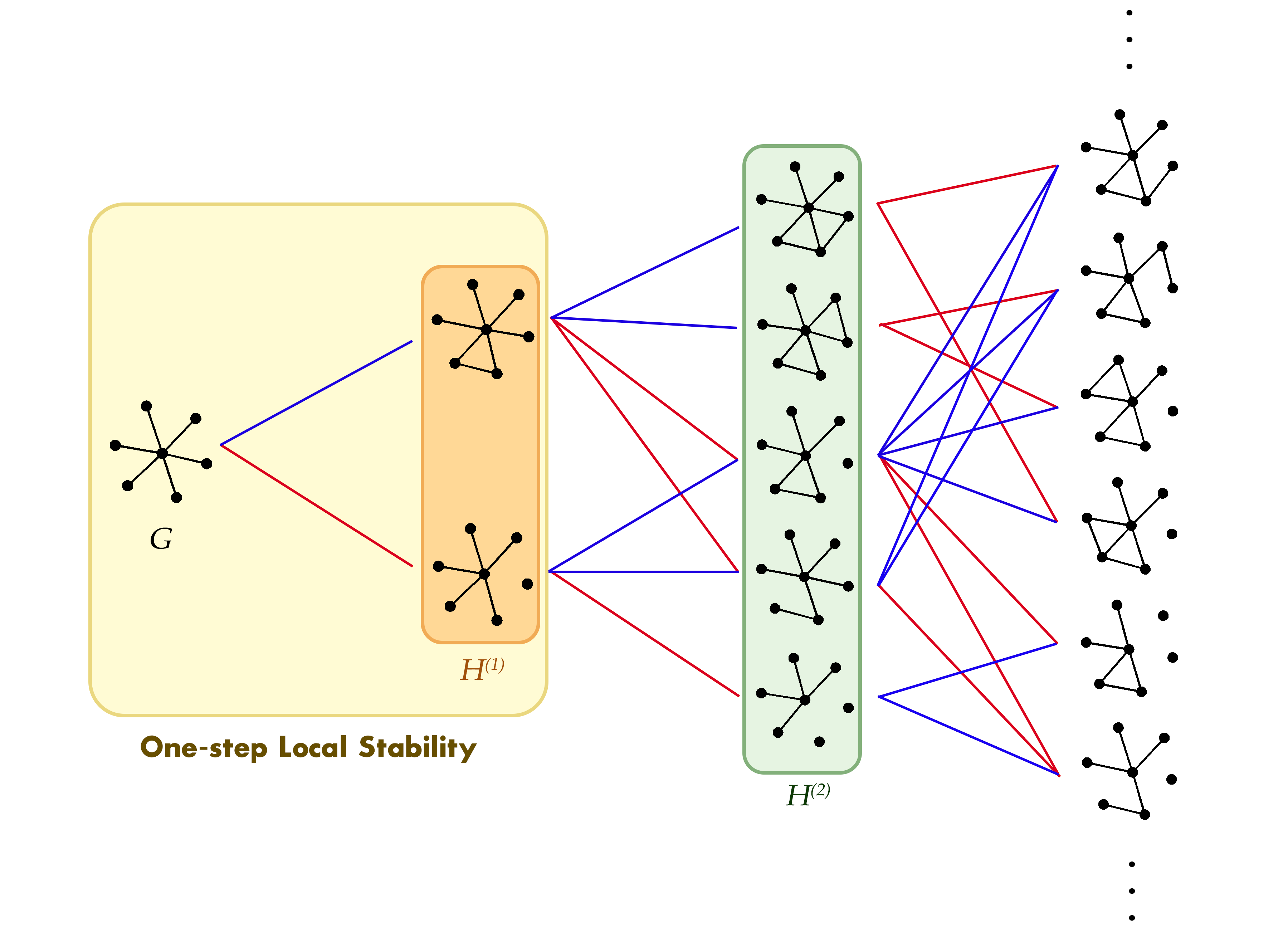}
  \caption{Examples of Hamming trajectories that could be taken from a target graph $G$. Here we group all isomorphic graphs into the same representation, and we only consider trajectories without loops. Each step is shown as either a blue line or a red line. Blue lines indicate steps involving edge addition and red lines indicate edge deletion. In this example, the set $H^{(1)}$ only contains two graphs, and local stability is satisfied if moves to either of the two graphs are disfavored.}
  \label{fig:traj}
\end{figure}

\begin{figure}[h!]
  \centering
  \includegraphics[width=10cm]{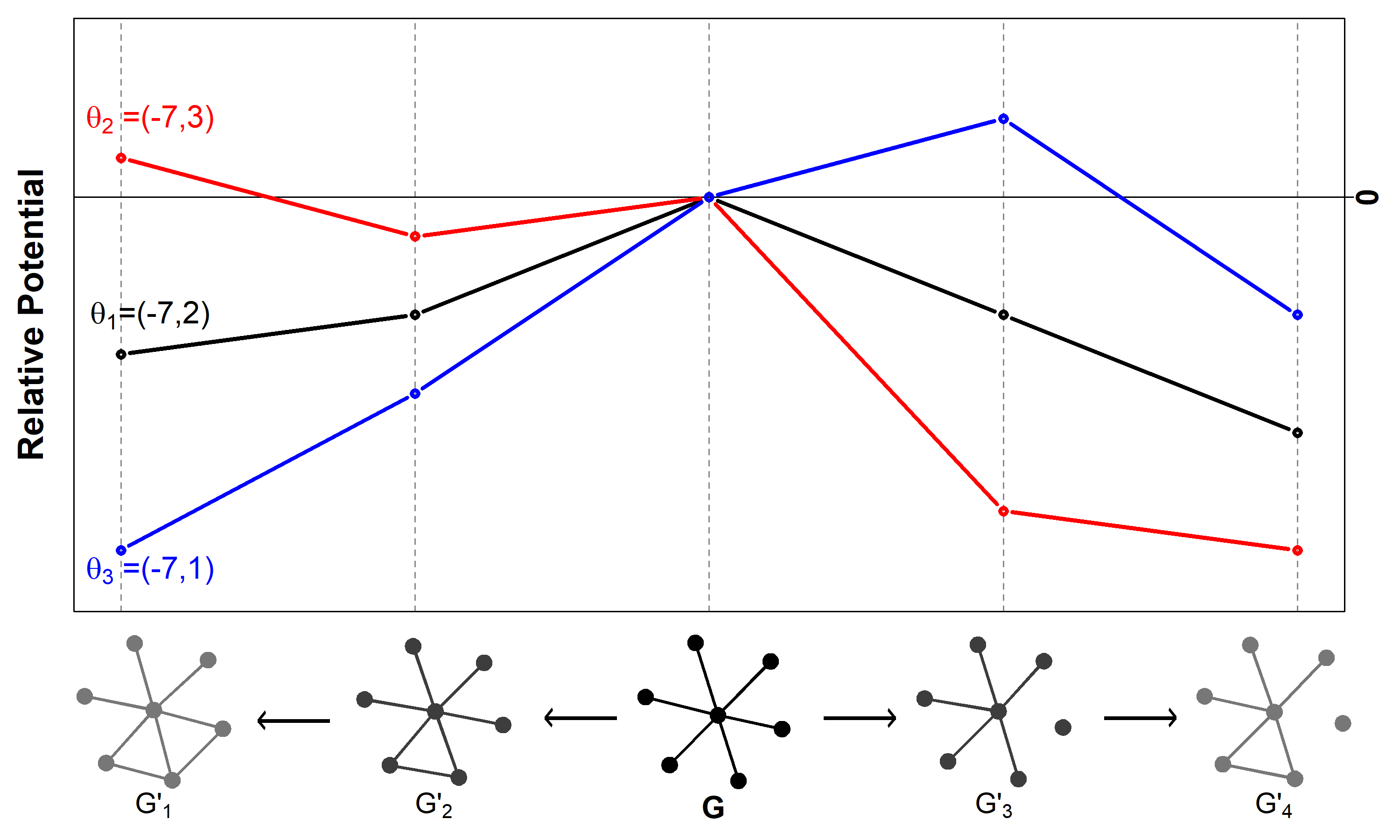}
    \caption{Illustration of a local mode in Hamming space. We choose two example trajectories, both started from the target graph $G$ (plotted in the center). Trajectory $G$-$G'_2$-$G'_1$ is plotted from $G$ to left and trajectory $G$-$G'_3$-$G'_4$ is plotted from $G$ to right. Relative potential is the potential gain moving away from $G$. The model here has two terms: \texttt{edges} and \texttt{kstar(2)}. We choose three different $\theta$'s, and the relative potentials for each choice of $\theta$ are plotted in different colors. Under the model with coefficients $\theta_1$ (in black), $G$ has highest potential among these five graphs. In fact, $G$ has the highest potential among all graphs in $H^{(1)}$ and $H^{(2)}$, indicating that $G$ is stable against any two-step toggles. Under $\theta_2$ (in red), $G$ is \emph{locally stable} because it has higher potential than both graphs in $H^{(1)}$. Note that although $G'_1$ has higher potential than $G$ under the coefficients $\theta_2$, trajectory to $G'_1$ is separated by graph $G'_2$ whose potential is lower than $G$. Under the coefficients $\theta_3$ (in blue), $G$ is considered unstable, due to the fact move to $G'_3$ increase the potential. In fact, coefficient $\theta_3$ stabilizes $G'_3$ instead.} 
  \label{fig:traj_pot}
\end{figure}

The above stability definition can be applied to any arbitrary set of alternative graphs up to and including $S=\mathcal{G}\setminus G$; in the latter case, stability of $G$ under $\mtheta$ is obviously synonymous with $G$ being the mode of the model specified by $\mtheta$.  Typically, however, we are interested in evaluating the stability of $G$ with respect to a set of alternatives that are ``close'' to $G$ under a hypothetical change process (e.g., the stochastic choice process of Snijders \cite{snijders:sm:2001}, the reaction kinetics of Grazioli et al. \cite{grazioli.et.al:jpcB:2019}, or the Metropolis dynamics frequently used in MCMC algorithms \cite{hunter.et.al:jcgs:2012}) in which the network of interest evolves via discrete changes in which single edges are added or removed from the graph (\emph{dyad toggles}) such that ``uphill moves'' on the probability surface occur at a higher rate than ``downhill moves.''  Here, we discuss a specification of $S$ that is broadly useful for assessing local stability in such settings. The intuition is as follows.  Let $S$ be the set of all graphs in $\mathcal{G}$ reachable from $G$ by a single dyad toggle (i.e., the Hamming sphere of radius 1 centered on $G$).  When $G$ is stable with respect to $S$, moves away from $G$ will be disfavored (and return moves will be at least somewhat favorable); by contrast, when $G$ is unstable with respect to $S$, there will be favorable dyad toggles that move away from $G$ (with the return move being unfavorable).  This notion of local stability (equivalent to $G$ being a local mode in the Hamming space on $\mathcal{G}$) generalizes naturally to higher Hamming radii, and is a natural and easily computed starting point for considering longer trajectories (see fig.\ref{fig:traj} and fig.\ref{fig:traj_pot}).

Specifically, let $H^{(d)}$ be a set of all graphs that are Hamming distance $d$ away from the target graph $G$.  $G$ may be said to be locally stable at (Hamming) radius $d$ if $G$ is stable with respect to $S=H^{(d)}$.  In the special case of a single edge change, we are interested in $d=1$. Since the number of rows in $M$ is the cardinality of $S$, the stable cone for a simple graph with $v$ vertices arises from the intersection of $\tfrac{1}{2} v(v-1)$ half-spaces.  Although this quadratic scaling (more generally, $\mathcal{O}(v^{2d})$) is unfavorable, it is typically the case that symmetries associated with $\mt$ lead many rows of $M$ to be identical; such redundant rows can be removed without changing the solution to the stability problem.  In some cases, the resulting compression can be considerable: for instance, one of the examples we show in section~\ref{sec:star_stability} leads to a two-row $M$ matrix regardless of $v$.  Moreover, some unique rows of $M$ may also be redundant, in that they specify constraints on the stable region that are weaker than the constraints imposed by the other rows of $M$.  As we show below, this form of redundancy can be exploited to calculate the stable cone in practical settings (as illustrated in section~\ref{sec:lazega}).




\subsection{Solving for the Stable Cone}

We employ the following terminology to refer to geometric features associated with the stability problem. Assuming we are in a $K$-dimensional parameter space, a $K-1$-dimensional dividing hyperplane separates the region of the parameter space where the target graph is more probable than a specific alternative graph $G'\in S$ from the region of the parameter space where the target graph is at most equiprobable to $G'$.  A dividing hyperplane may or may not constrain the stable cone (the intersection of half-spaces imposed by all dividing hyperplanes associated with the rows of $M$).   When it does, the ``side'' of the cone created by its intersection with the hyperplane is called a \emph{facet}. An intersection of two facets is called a \emph{ridge}, which is an element of dimension $K-2$. 

\begin{figure}[h!]
\centering
  \includegraphics[width=5cm]{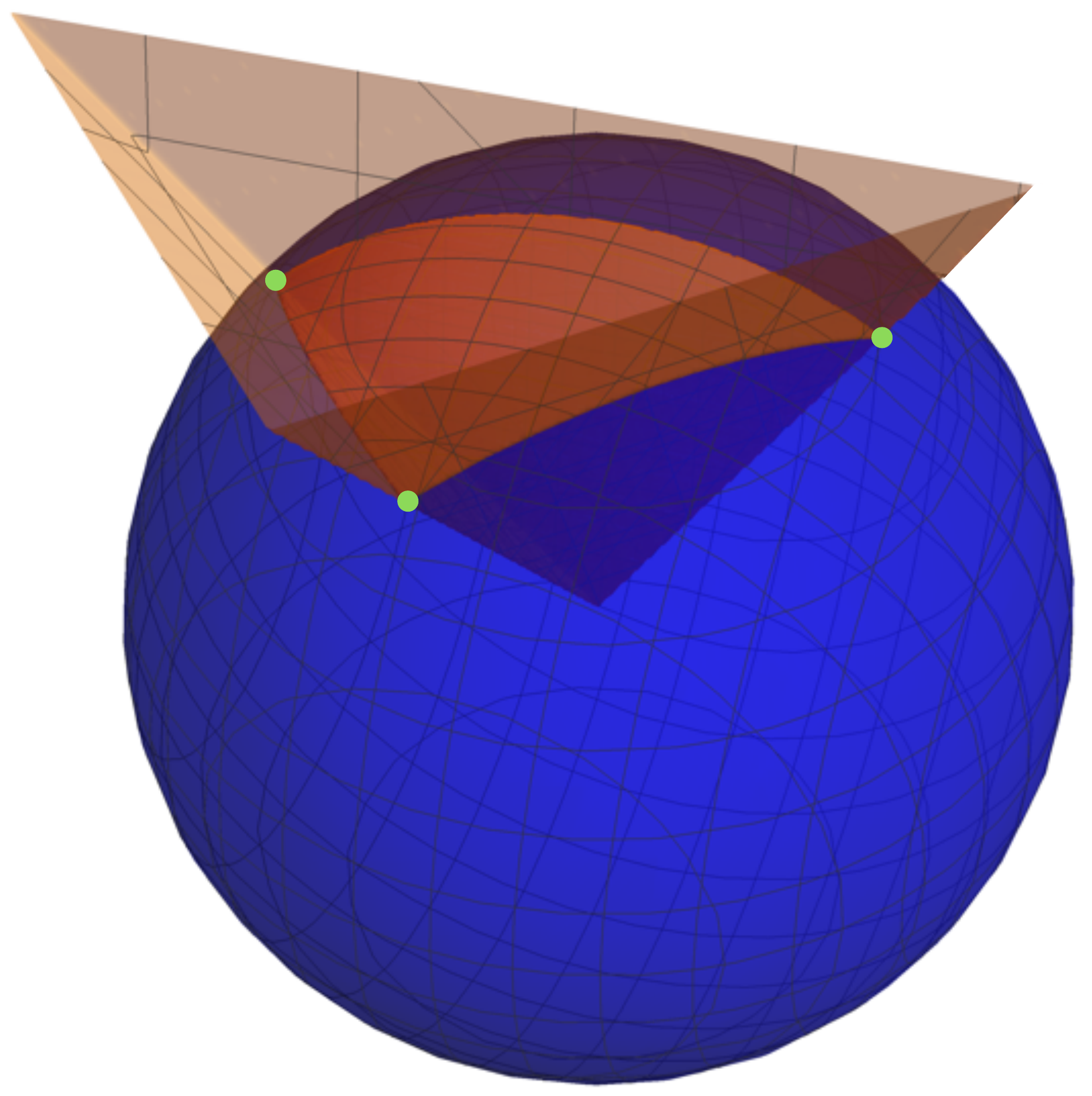}
 \caption{The vertex representation (intersections of the transparent orange planes) can be conveniently stored as the points (green dots) produced by the intersection of the vertices with a hypersphere (blue sphere)}
  \label{fig:hyper}
\end{figure}

The stable cone can be characterized using two different geometric representations: the \emph{halfspace representation} (or \emph{H-representation}) and the \emph{vertex representation} (or \emph{V-representation}) \cite{avis1992pivoting}. The H-representation characterizes the subspace $\Phi$ as a set of conditional linear inequalities, or half-spaces, as shown by the blue facets in Fig. \ref{fig:convex_cone}, and in Eq. \ref{hRepEq} below:
\begin{equation}
\Phi = (\theta | M_i\theta < 0)  \text{   } \forall i \in \{1, 2, ..., k\}
\label{hRepEq}
\end{equation}

\noindent while the V-representation characterizes the subspace as the convex hull generated by the vertices that are created by each pair of intersecting planes from the H-representation (shown as black arrows in Fig. \ref{fig:convex_cone}). It is worth noting that, while solving for the stable region, it is convenient to store each vertex as a single point in parameter space by calculating the intersection of each vertex with a hypersphere of a given radius (shown in Fig. \ref{fig:hyper}). Since all points along the ray associated with each vertex can be obtained by rescaling, as can the ray associated with any point within this particular slice of the stable region (solid orange triangle in \ref{fig:hyper}), no information is lost by using this representation. Additionally, the data structure for the V-representation is such that the indices of the hyperplanes whose intersection comprise each vertex is also stored in the V-representation object. The remainder of this section presents the calculation of the stable region using this method for storing the V-representation. 

\begin{figure}[h!]
\centering
  \includegraphics[width=16cm]{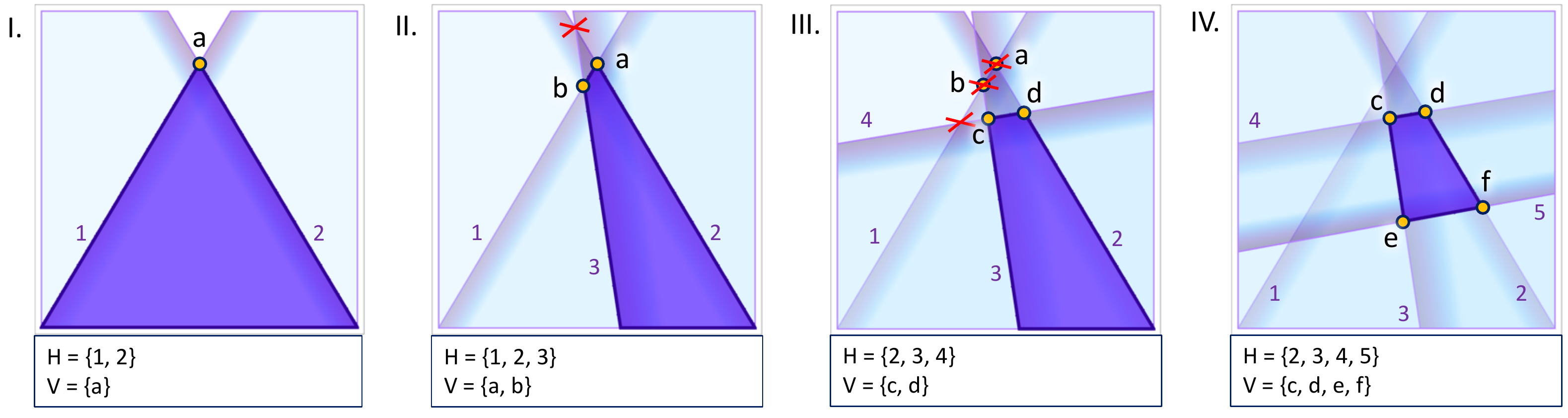}
 \caption{A demonstration of Algorithm I, used to define an initial closed superset of the stable region. I.) The first step in Algorithm I: the H-representation is initialized with two non-parallel halfspaces drawn from \textbf{M}, while the V-representation is initialized as the intersection of the two halfspaces in H. II.) Step 2 in algorithm I: A third halfspace is introduced, all n choose 2 intersections are calculated, all intersections within the stable region defined by \emph{H} become the V-representation, and all halfspaces whose intersections comprise \emph{V} become \emph{H}. The convex hull is not closed, so we iterate another step. III.) Step 3 in algorithm I: as in the previous step, a new halfspace is introduced, stable intersections become \emph{V}, and their respective halfspaces become \emph{H}. Note that in this case, two previously included vertices (a and b) are now excluded from \emph{V}, as is halfspace 1, since it no longer contributes any stable intersections to \emph{V}. Still, the convex hull is not closed, so we iterate another step. IV.) A new halfspace is introduced, and the process from the previous two steps is repeated. This time, the convex hull is closed, thus Algorithm \ref{InitStableAlgo} is terminated, and an initial closed superset of the stable is returned.}
  \label{fig:algo1}
\end{figure}

\begin{figure}[h!]
\centering
  \includegraphics[width=12cm]{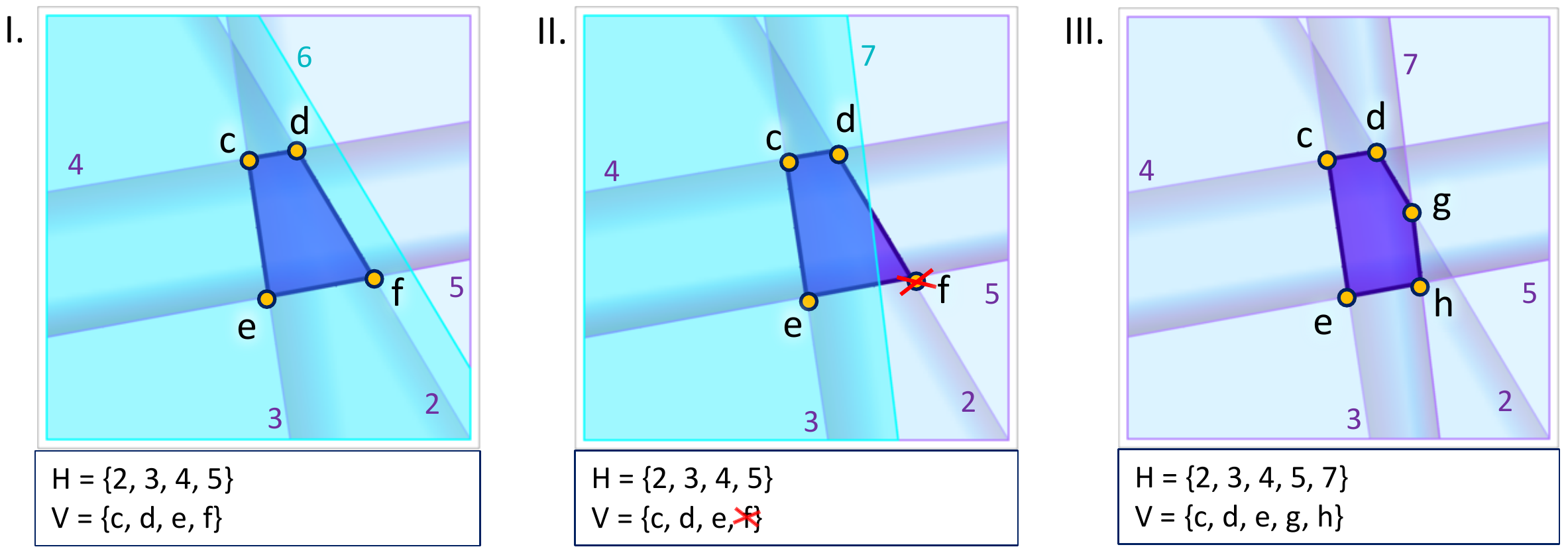}
 \caption{A demonstration of Algorithm II, which takes an initial closed superset of the stable region and an \textbf{M} matrix as input, and outputs \emph{H} and \emph{V} descriptions of the stable region for a given model. I.) The closed set from figure \ref{fig:algo1} (step IV) becomes the input for Algorithm II.) A new halfspace is introduced (cyan), but since no vertices from \emph{V} are excluded, the halfspace is redundant, and thus rejected. II. Another halfspace is introduced, which excludes vertex \emph{f}, thus \emph{f} must be rejected from \emph{V} and all intersections between the new halfspace and those whose intersections comprise \emph{f} must be calculated and tested for stability under \emph{H}. III.) The new halfspace is appended to \emph{H} and the intersections between 2, 5, and 7, stable under \emph{H} are appended to \emph{V}.}
  \label{fig:algo2}
\end{figure}

Although the H and V representations each offer a complete description of the stable cone individually, they are perhaps most useful for solving the stable cone when the two representations are used in tandem. In our application, we leverage the complementarity of the two representations to iteratively search for both a non-redundant H-representation and V-representation simultaneously by applying a methodology based on the \emph{double description method} (or \emph{DD method}), introduced by Avis \cite{avis1992pivoting}. Given an initial redundant \textbf{M} matrix (\emph{redundantHalfspaces}), with all edge changes of interest represented in the rows, the stable cone for a given target graph and set of sufficient statistics can be solved by first using Algorithm \ref{InitStableAlgo} (Fig. \ref{fig:algo1}) to obtain an initial closed superset of the stable region, which is then passed (along with \textbf{M}) to Algorithm \ref{DDalgo} (Fig. \ref{fig:algo2}), where the initial closed superset is whittled down to a non-redundant double description of the stable region. 

Given the redundant \textbf{M} matrix for a target graph and a set of sufficient statistics, this information is first passed to Algorithm \ref{InitStableAlgo} in order to calculate an initial closed superset of the stable region. Algorithm \ref{InitStableAlgo} begins by first obtaining a set of $|\theta| - 1$ rows from \textbf{M}, where no two halfspaces are parallel, and calculating their intersection with both each other and the hypersphere. These rows and this vertex point serve to initialize the H-representation ($H$) and V-representation ($V$), respectively. At this point, a new row is drawn from \textbf{M} and appended to $H$, and intersections between all possible combinations of $|\theta| - 1$ rows of \textbf{M} are calculated and then appended to $V$. Although the current iteration of $H$ is redundant, its stable region is still equivalent to the non-redundant form, thus $H$ is used to remove all unstable vertices from $V$. Given that the data structure for $V$ includes the labels for the rows of $H$ that intersect to form each vertex in $V$, it is then trivial to now remove all rows of $H$ that are not represented in $V$. At this point, $H$ and $V$ are both non-redundant. The final step in the loop is to check the convex hull for closure. If the convex hull represented by $H$ and $V$ is closed, Algorithm \ref{InitStableAlgo} returns $H$ and $V$ and terminates. One straightforward method for testing the closure of the convex hull described by $V$ is to first check that the number of vertices is $\geq |\theta|$, and then use any of the many available methods for calculating convex hulls from points (e.g. Quickhull \cite{barber1996quickhull}) to calculate the convex hull of the vertices in $V$. If both the number of vertices is $\geq |\theta|$, and the number of halfspaces in the halfspace representation returned by the convex hull finding method is equivalent to the number of rows in $H$, the stable region represented by $H$ and $V$ is closed. This test for closure is made possible by the fact that if the convex hull finding algorithm is operating on a convex hull that is open with respect to the stable region, it will introduce a new halfspace to close off the open end of the space.

~
  
\begin{algorithm}[H]
 \label{InitStableAlgo}
 \SetKwProg{Fn}{Function}{}{}
\caption{Finding an initial closed superset of the stable region}
\KwData{\textbf{M}}
\SetAlgoLined
 initialize H; V\;
 bool are.parallel = TRUE\;
 \While{are.parallel}{
 h.init = sample.two.rows(\textbf{M})\;
 are.parallel = check.parallel(h.init)
 }
 
 H = h.init\;
 V = get.exhaustive.intersections(H)\;
 bool hull.is.closed = FALSE\;
 \While{!hull.is.closed}{
     h.test = sample.one.row(\textbf{M})\;
    H = append(H, h.test)\;
    vertices.new = get.exhaustive.intersections(H, h.test)\;
    V = append(V, vertices.new)\;
    V = return.stable.vertices(V, H)\;
    H = get.H.from.V(V)\;
    hull.is.closed = testForClosedConvexHull(V)
 }
 
 \Return \{H, V\}
 \end{algorithm}

Once Algorithm \ref{InitStableAlgo} has returned an initial closed superset of the stable region, $H$, $V$, and \textbf{M} are then passed to Algorithm \ref{DDalgo}. The premise of Algorithm \ref{DDalgo}, as put forth by \cite{avis1992pivoting}, is that given an initial closed convex hull, any newly introduced halfspaces are only non-redundant if their introduction excludes one ore more previously existing vertices. The algorithm also leverages the fact that the data structure for $V$ keeps track of which halfspaces intersect to produce each vertex, in that upon introduction of a new non-redundant halfspace, only the halfspaces whose intersections comprise the newly excluded vertex or vertices must be included in the calculation of newly created vertices. Intuition for the methodology can be readily obtained from Fig. \ref{fig:algo2}.

 \begin{algorithm}[H]
 \label{DDalgo}
 \SetKwProg{Fn}{Function}{}{}
\caption{Using the DD method to solve the stable cone}
\KwData{H.initial, V.initial, \textbf{M}}
\SetAlgoLined
 initialize H = H.initial; V = V.initial\;
 
 \For{all rows in \textbf{M}}{
     h.test = get.next.row(\textbf{M})\;
    v.excluded = get.excluded.vertices(V, h.test)\;
    \uIf{is.not.empty(v.excluded)}{
    H = append.row(H, h.test)\;
    rows.for.testing = get.rows.comprising.vertices(v.excluded)\;
    vertices.for.testing = get.new.intersections(rows.for.testing, h.test)\;
    v.new = get.stable.vertices(vertices.for.testing)\;
    V = append(V, v.new)
    }
 }
\Return \{H, V\}
 \end{algorithm}

The order of computational complexity for this methodology is best understood using a bounding argument. First, we establish the two fundamental operations employed by our methodology: 1) calculating intersections of sets of halfspaces that form vertices, and 2) testing whether or not vertices lie within the stable region. The first operation is an inversion of a $d \times d$ matrix, where $d$ is the dimensionality of the model (i.e. the number of sufficient statistics), and the second operation is a simple matrix multiplication of a vector of length $d$ by the matrix representing the current iteration of the H-representation (with trivial parallelization); thus, the rate-limiting operation is the calculation of vertices. Next we establish the scaling for a brute force treatment whereby, first, all $n$ choose $d-1$ intersections between hyperplanes are taken to obtain all possible vertices (operation 1), then each vertex is tested to determine whether or not it is within the stable region (operation 2), leaving the non-redundant double description. In this case, the computational complexity scales as $\mathcal{O}((\frac{en}{d})^d)$. The worst-case scenario for our methodology would be that \emph{Algorithm 1} fails to produce a closed stable region until the very last row from the $M$ matrix has been included in the exhaustive search, demonstrating that, at its worst, the order of complexity for our algorithm is equal to the exhaustive brute force treatment. Next, consider the best-case scenario: given that a particular stable region can be expressed as an H-representation comprised of $h$ rows, suppose that these are the first $h$ rows selected from the $M$ matrix in \emph{Algorithm 1}. In this case, the only vertices calculated would be those produced by the H-representation implying a computational complexity of $\mathcal{O}((\frac{h}{d})^d)$. Even for a pessimistic case where only half of the rows in $M$ are redundant ($h = nrow(M)/2$), our methodology would offer a computational speedup of $.5^{-d}$, a massive speedup over the fully exhaustive brute force method (e.g. $32\times$ speedup for a network model with just 5 sufficient statistics).


\section{Case Study I: Cult (Star) Structure}
\label{sec:starStructure}
In this section, we apply the above method to a simple example of a social structure that we here refer to as the \emph{cult network}. A cult network is characterized by an isolated star structure, where there is a single core (``leader'') node in the center connecting to all peripheral (``follower'') nodes, and no peripheral nodes are mutually adjacent. Thus, a follower can reach any other follower, but only via a path that is brokered by the leader. Although the motivation behind introducing our methodology using a cult network as an example is illustrative simplicity, it does represent a stylized version of features found in some real-world charismatic cults. For example, both the People's Temple \cite{johnson:sa:1979} and Heaven's Gate \cite{davis:jaer:2000} cults featured leaders who eventually isolated their groups from outside contact and regulated the flow of critical information such that rank-and-file members were encouraged to trust and obey only the leaders themselves (with unmediated intra-group relationships among members being heavily discouraged).  Maintaining these star-like group structures required considerable creativity and effort on the parts of the cult leaders, who ultimately crafted complex social environments that stabilized what would otherwise be highly unfavorable pattern of social relationships.  The cult network model presented here demonstrates the conditions that are necessary for such a structure to be stabilized under one very simple class of social processes.

The model family employed here is by design minimal, incorporating only two types of social ``forces:''  a general propensity to form edges, and a force that governs the propensity for untied pairs of individuals to have or lack partners in common.  In ERGM form, such a family corresponds to a vector of statistics ($\mt$) containing respectively the count of edges (\texttt{edges}) and the count of null dyads having no partners in common (\texttt{nsp(0)}).  We investigate whether the star structure can be stabilized under this model family, and if so, define the region of the parameter space where that the network is stable. 

In the remainder of this section, we first compute the stable cone, then demonstrate that this region aligns with stability as assessed by simulated Metropolis dynamics. We also consider the question of which dyad is likely to be the first to be toggled (assuming that some toggle occurs).  This approach can help identify which edges/nulls are most vulnerable to change under the model. We can also show analytically the probability of the target network transitioning to each of the alternative networks, given that an edge change has occurred. We then validate these results with simulated trajectories. 

\begin{figure}[h!]
\centering
  \includegraphics[width=9cm]{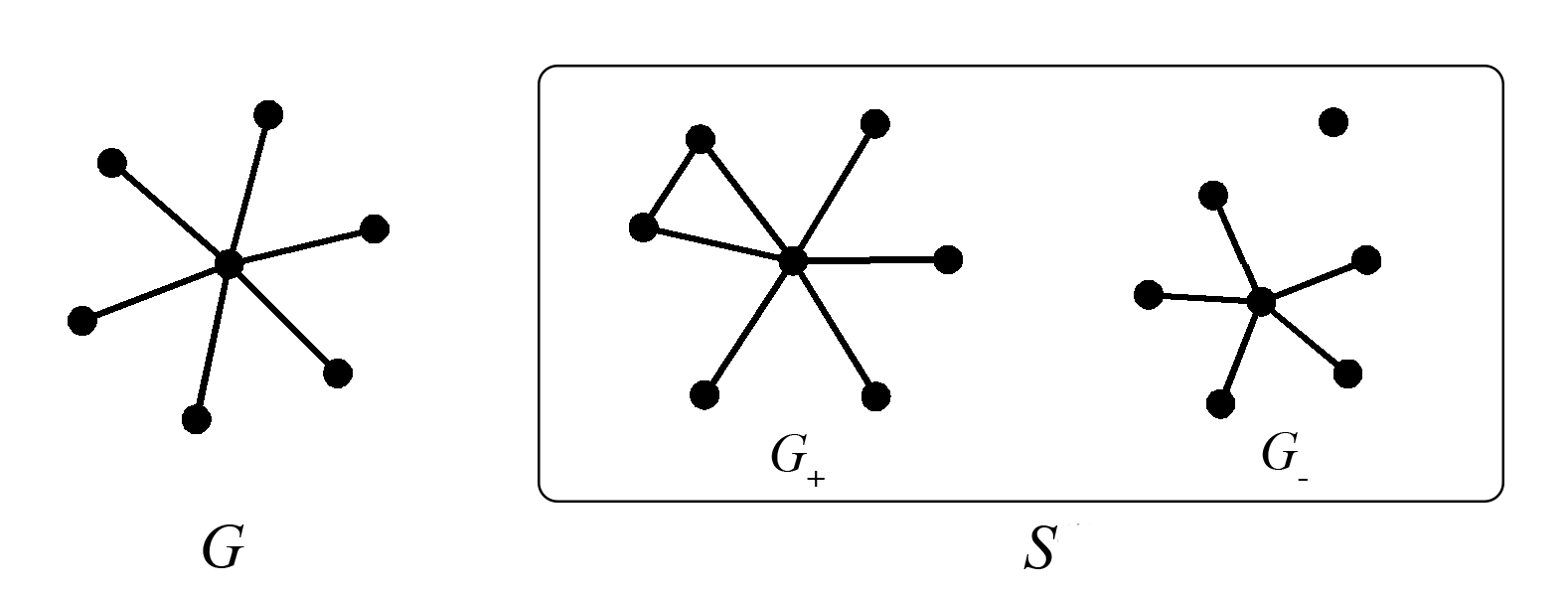}
  \caption{$S=H^{(1)}$ for the star structure.}
  \label{fig:star-ex}
\end{figure}

We use an undirected star network of $v$ vertices as the target network. For illustration purpose, we let $v = 7$ when figures or numeric results are generated (e.g., $G$ in figure \ref{fig:star-ex}). For simplicity, we assume that no attribute is associated with vertices or edges, indicating that all peripheral nodes are interchangeable. Because our model is unable to differentiate between the target graph and other members of its isomorphism class, all isomorphic graphs will be counted towards stability.  The set $S =H^{(1)}$ contains only two distinct networks (illustrated in fig.~\ref{fig:star-ex}): $G_+$ is the result of adding one edge to $G$, connecting two peripheral nodes; and $G_-$ is the result of removing one edge from $G$, breaking one connection between the core and a peripheral node. As mentioned, the two sufficient statistics are \texttt{edges} and \texttt{nsp(0)}).  We note in passing that there are other model families that can also generate star structures.  For instance, a term influencing the number of null dyads having exactly one shared partner (\texttt{nsp(1)}) can also be used to generate star-like graphs. However, the \texttt{nsp(0)} term, despite being less obviously associated with the star structure, interacts with the \texttt{edges} term to anchor the structure in place, i.e. a negative value in \texttt{nsp(0)} suppresses the existence of non-edges sharing exactly zero partners, thus non-edge pairs with higher shared partners are boosted, to a limit that is controlled by the network density (set by \texttt{edges}). 

\subsection{Stable Parameter Region of the Star Structure}
\label{sec:star_stability}

The rows of the $M$ matrix are comprised of the change scores (i.e., differences of graph statistics) between the target graph and the graphs in the alternative set (shown in table~\ref{table:stats}). The parameter space of this model is 2-dimensional - each sufficient statistic of the model occupies one dimension. To illustrate the parameter space we plot \texttt{edges} on the horizontal dimension ($x$-axis) and \texttt{nsp(0)} on the vertical dimension ($y$-axis). For simplicity, hereafter we use $x,y$ to represent the parameter values of \texttt{edges} and \texttt{nsp(0)}, respectively.

\begin{table}[ht]
\centering
\begin{tabular}{l cc c cc}
  \hline
  &\multicolumn{2}{c}{ Sufficient Statistics ($t$)}  && \multicolumn{2}{c}{Change Score ($M$)} \\
 & \texttt{edges} & \texttt{nsp(0)} && \texttt{edges} & \texttt{nsp(0)} \\ 
  \hline
$G$ & $v$-1 & 0 && - & - \\ 
\ga & $v$ & 0 && 1& 0 \\ 
\gd & $v$-2 & $v$-1 && -1 & $v$-1\\ 
   \hline
\end{tabular}
\caption{Sufficient statistics and change scores for the star structure $G$ and the set $S=H^{(1)}$. $v$ is the number of vertices in the graph; \texttt{edges} and \texttt{nsp(0)} are the two ERGM terms used in the model.}
\label{table:stats}
\end{table}

\begin{figure}[h!]
\centering
  \includegraphics[width=10cm]{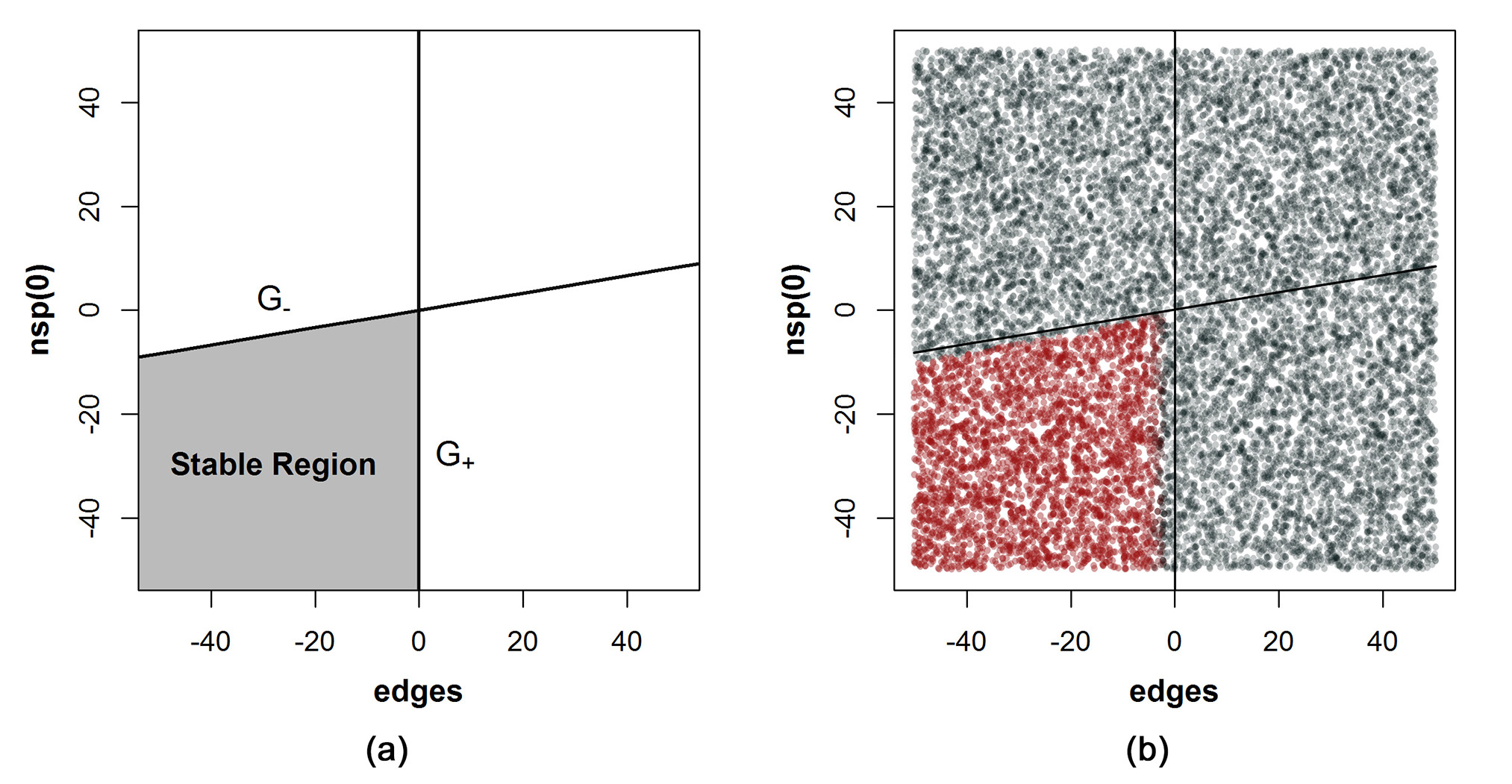}
  \caption{\textbf{a}. The stable cone (grey region) for the star structure under the \texttt{edges}/\texttt{nsp0} family, assuming $v=7$. \textbf{b}. Fraction of simulated networks remaining in star configurations after $10^6$ random-walk Metropolis steps for selected parameter values (dots).  Color values range from grey (no persistence) to red (complete persistence).  Despite being defined only relative to $H^{(1)}$, the stable cone (line-bounded region in lower left) closely matches the region of long-term dynamic stability.}
  \label{fig:star_stable_cone}
\end{figure}

 
As derived in section~\ref{sec:propertisOfSR}, each graph in $S$ defines a half-plane (or more generally, a half-space; in subsequent discussions, we use general terms such as half-spaces and hyperplanes, despite them being called half-planes and lines in 2-dimensional space): $G+$ defines $x\leq0$, and $G_-$ defines $x-(v-1)y\geq0$. The stable region, which is the intersection of the two half-spaces, is plotted as the shaded area in fig.~\ref{fig:star_stable_cone}a (assume $v=7$). As proved in section~\ref{sec:propertisOfSR}, the stable region is a cone that has infinite height and points at the origin. 

To verify that the stable cone is compatible with stability under explicit dynamics, we sample $10^6$ parameter vectors and run 50 random walk Markov chains at each sampled vector.  (All simulation was performed using the \texttt{ergm} package within \texttt{statnet} \cite{hunter.et.al:jss:2008,handcock.et.al:jss:2008} using random dyad proposals.)  Each chain starts from the perfect star structure and runs for $t = 10^6$ Metropolis steps. We examine whether the star structure is preserved in the networks returned at the end of the simulation, and plot the fraction of the networks that remain unchanged (fig.~\ref{fig:star_stable_cone}b).


Even though the stable cone is based only on stability versus $H^{(1)}$, it corresponds closely to long-term stability under Metropolis dynamics (red points in Fig.~\ref{fig:star_stable_cone}b). Outside of the stable cone, the graph is dynamically unstable (dark grey points in fig.~\ref{fig:star_stable_cone}b Along the interior of the two facets, there is a thin stripe of models that are dynamically unstable despite being inside the locally stable region. This is reflective of the fact that, while moves away from $G$ are unfavorable, they will eventually happen given enough attempts (here, $10^6$).  Parameter values close to the faces of the stable cone have a lower margin of stability, in the sense that probability gap between $G$ and the elements of $S$ is smaller, and dynamic stability hence begins to be lost as one moves from the center of the locally stable cone to its faces.  An interesting observation is that the ``unstable band'' associated with $G_+$ is a bit wider compared with the band associated with $G_-$. This phenomenon can be explained by examining the ERGM potential of each alternative graph.

\begin{figure}[h!]
\centering
  \includegraphics[width=12cm]{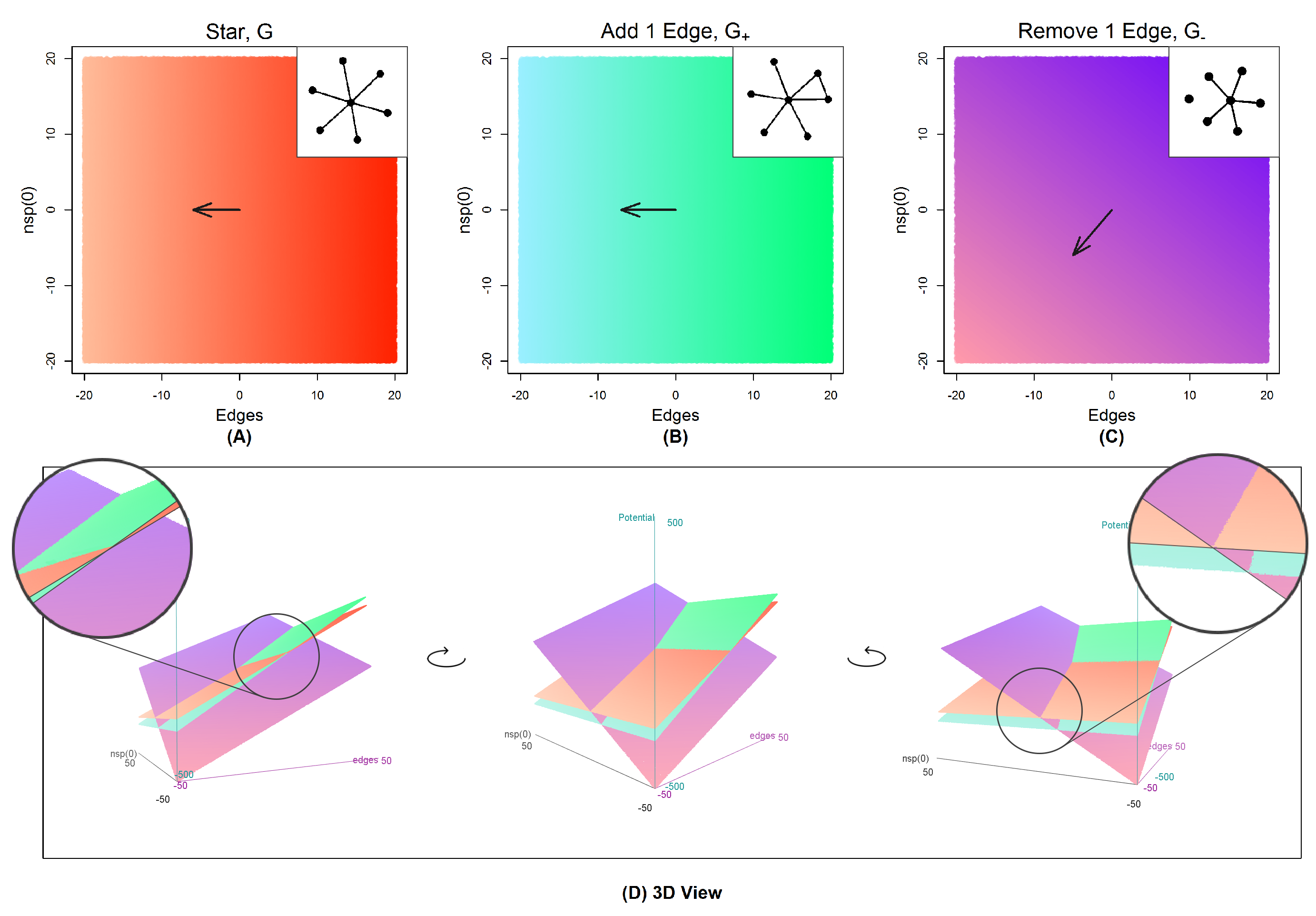}
  \caption{Potential planes of target graph $G$, and both alternative graphs $G_+, G_-$ in $S$. Panels A, B, and C plot the potential of networks $G$, $G_+$ ,and $G_-$, respectively. The gradients of the potential surfaces are plotted as arrows. Panel D plots the three gradients onto the same space. Note that the stable cone can be identified in the bottom left quadrant, where the potential of $G$ is higher than both $G_+$ and $G_-$ (see inset).}
  \label{fig:star_gradient-planes}
\end{figure}

By definition, the stable cone is the region where the ERGM potential of the target graph, $\mtheta^T\mt(G)$, is greater than any of the other graphs in $S$; facets and ridges of the cone then correspond to the sets of $\mtheta$ values where at least one or two graphs (respectively) in $S$ have the same potential as the target graph. Because the graph probability is proportional to the exponentiated ERGM potential, $\Pr(g) \propto \exp(\mtheta^T\mt(g))$, examining graph potentials provides insight into the behavior of the Metropolis dynamics. As shown in fig.~\ref{fig:star_gradient-planes}, the potential difference of $G$ and \ga is much more gradual along the dividing hyperplane comparing with that of $G$ and \gd, thus the down-potential step is taken with higher probability during the Markov Chain run. 

\begin{figure}[h!]
  \centering
  \includegraphics[width=10cm]{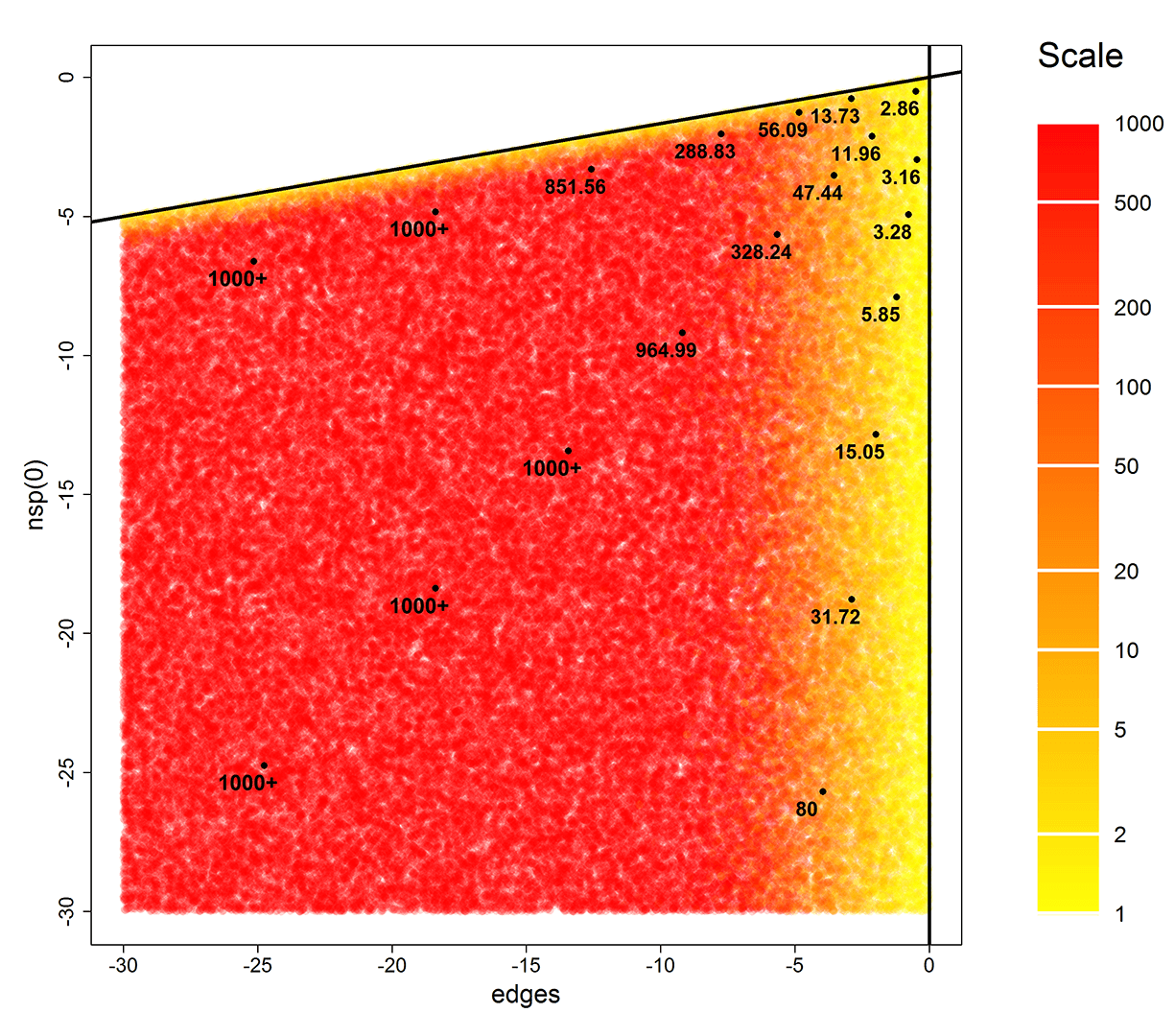}
  \caption{Mean number of Metropolis steps required for the first change in $G$, as a function of $\mtheta$.  Waiting times grow exponentially fast as one moves away from the faces of the stable cone.}
  \label{fig:star_stickiness}
\end{figure}

The above simulation experiment considers whether network snapshots under a long simulation are isomorphic to the target graph. In principle, a network could alternate between a number of states during the simulation, counting towards stability so long as it is in the target state when observed.  A more rigorous test of dynamic stability is to examine the expected time to the \emph{first} change under Metropolis dynamics across the stable cone.  Fig.~\ref{fig:star_stickiness} shows the mean number of steps required for the first accepted toggle, from yellow (1 step) to red ($>1000$ steps).  As in the previous simulation, we see that the graph strongly resists perturbation within the bulk of the stable cone, staying unchanged for more than 1000 simulation steps.  The wider stripe of dynamic instability along the $x = 0$ hyperplane ($P(G) = P(G_+)$) is associated with more rapid network changes, with fewer than ten simulation steps required on average as one nears the boundary of the stable cone.  The relatively sharp transition from the less dynamically stable region to strong dynamic stability results from the exponential decline in acceptance probabilities with respect to the potential difference: while our definition of $S$ leads to an extremely local definition of stability, the exponential decline in transition probabilities as one moves away from the faces of the stable cone makes it a good proxy for dynamic stability more broadly defined.

\subsection{Dyad Vulnerability}
\label{sec:star_vulnerable}

We have shown that the target structure is most dynamically stable well within confines of the locally stable cone, with dynamic stability weakening in the region closer to the facets of the cone.  Of course, given enough attempts, a change will inevitably occur; this raises the question of which structure in $S$ will be selected when the first dyad toggle is accepted.  This can be thought of as a problem of \emph{dyad vulnerability}: if change happens, which dyads are most vulnerable to being toggled?  In the case of the cult network, symmetry leaves us only two types of possible dyad toggles: toggles that break connections between the center vertex and one of the periphery vertices (denoted $d_-$,  resulting in a graph isomorphic to $G_-$), or toggles that establish connections among periphery vertices (denoted $d_+$, resulting in a graph isomorphic to $G_+$). 



Assume a dyad toggle $d_i$ is selected randomly from all possible $\tfrac{1}{2}v(v-1)$ toggles. The two types of toggles are proposed at rates based on the frequencies they appear in $G$: $\Pr(d_-) = \tfrac{2}{v}$, and $\Pr(d_+) =\frac{v-2}{v}$. Then the acceptance ratio ($ \alpha = \Pr(G_i)/\Pr(G)$) is calculated. The proposed toggle is accepted if the new graph is more probable, i.e., $\alpha > 1$. Otherwise, the toggle is accepted with probability $\Pr(\mbox{accept}|d_i) = \alpha =  \exp((t(G_i)-t(G)\theta^T)$. The acceptance probability if $d_-$ is proposed is:
\begin{equation*}
\Pr(\mbox{accept}|d_-, G) = 
\begin{cases}
     \exp(\mtheta(t(G_-) - t(G))),& \text{if }\mtheta(t(G_-) - t(G)) \leq 0\\
    1,              & \text{otherwise.}
\end{cases}
\end{equation*}


Then the probability of the target graph $G$ change to $G_-$ can be calculated as 
\begin{equation*}
\Pr(G_-|G) = \Pr(d_-)\Pr(\mbox{accept}|d_-).
\end{equation*}

Similarly, $\Pr(G_+|G)$ can be derived when $d_+$ is proposed. When proposed moves are rejected, then graph stay unchanged. Let $\mtheta =  [x,y]^T$ where $x$ is the parameter value for \texttt{edges} and $y$ is the parameter value for \texttt{nsp(0)}, the probability of becoming $G_+$, $G_-$, or stay unchanged given starting from the perfect star structure $G$ is:
\begin{align*}
\Pr(G_-|G)& = 
\begin{cases}
     \frac{2}{v}e^{-x+(v-1)y},& \text{if } x-(v-1)y \geq 0\\
     \frac{2}{v},  & \text{otherwise}
\end{cases},\\
\Pr(G_+|G)& = 
\begin{cases}
     \frac{v-2}{v}e^x,& \text{if } x \leq 0\\
     \frac{v-2}{v},  & \text{otherwise}
\end{cases},\\
\Pr(G|G)& = 1-\Pr(G_+|G)-\Pr(G_-|G)
\end{align*}
\begin{figure}[h!]
  \centering
  \includegraphics[width=9cm]{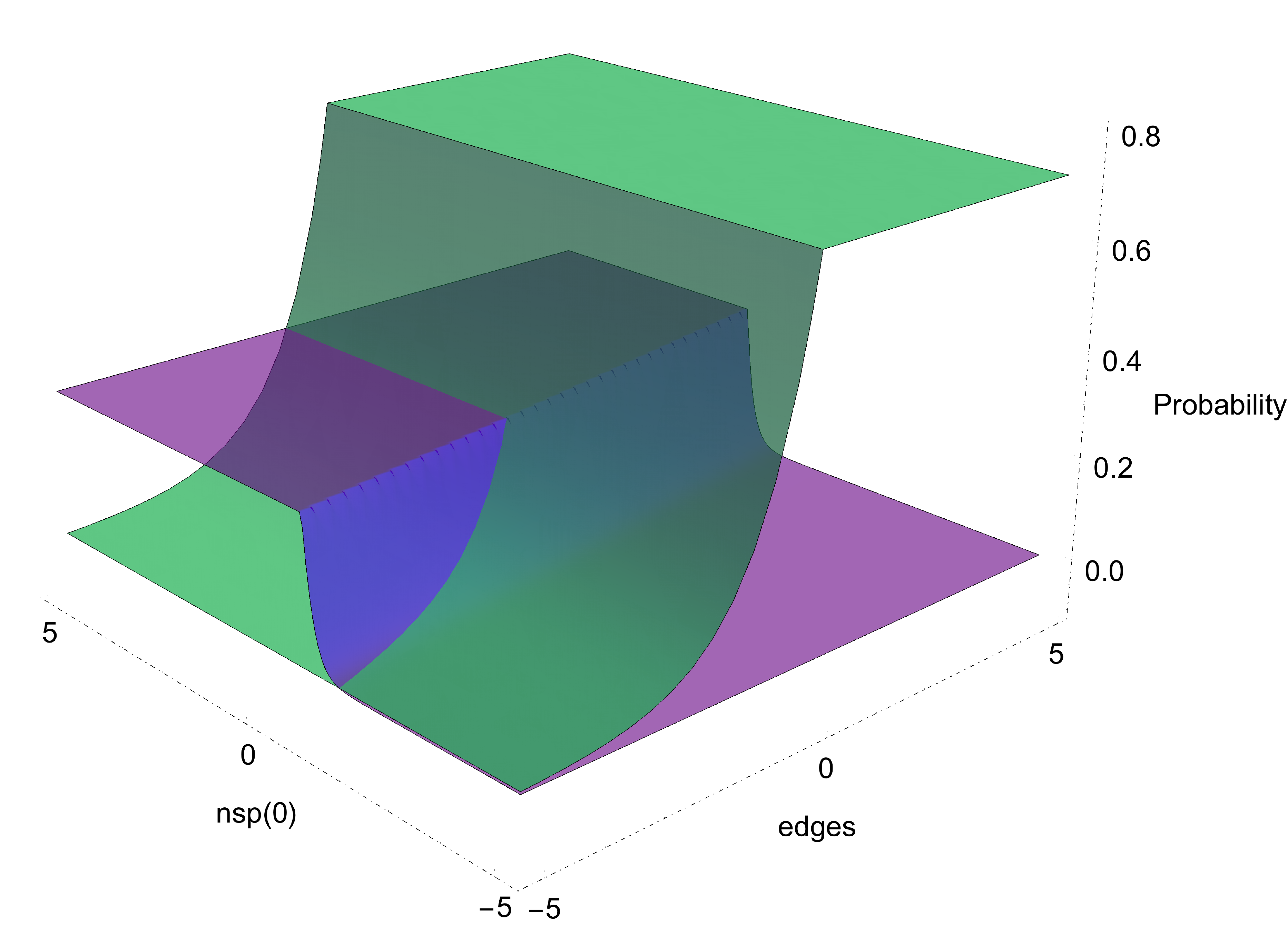}
  \caption{One step transition probability to each of the structures in $H^{(1)} = \{G_+, G_- \}$, as a function of the model parameters. The green surface corresponds to $G_+$, and the purple surface corresponds to $G_-$. The probability of staying unchanged as $G$ is 1 minus the sum of the two surface values.}
  \label{fig:dyad_vul_prob_surface}
\end{figure}

\begin{figure}[h!]
  \centering
  \includegraphics[width=7cm]{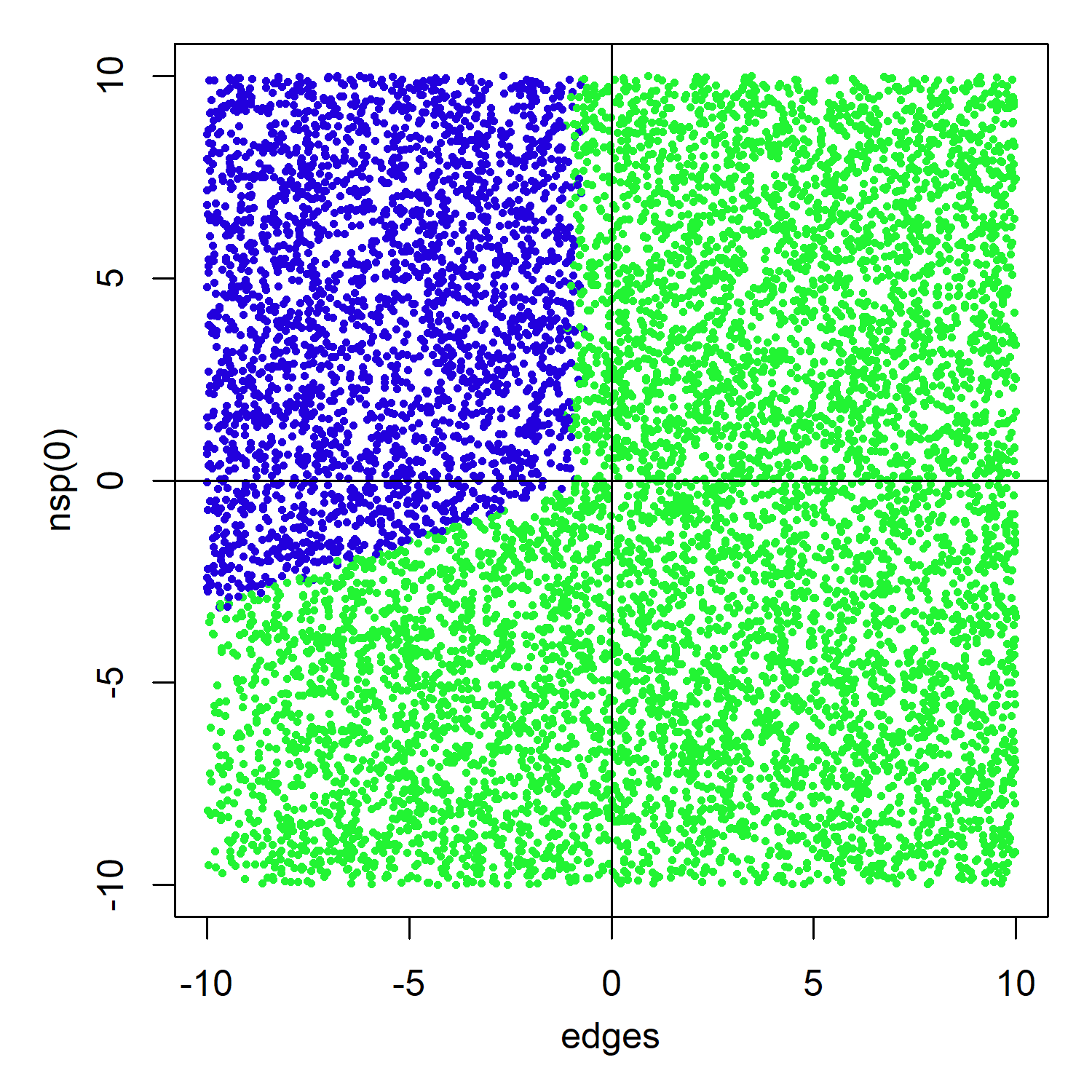}
  \caption{Simulation showing the most vulnerable dyad type, as a function of model parameters. Dots indicate sampled parameters, each of which was used to govern one MCMC trajectory; blue dots are networks that changed to $G_-$ on the first toggle, indicating that breaking an existing connection is more likely, and green dots are networks that first changed to $G_+$, indicating establishing a connection between two periphery nodes are more likely.}
  \label{fig:dyad_vul_simulation}
\end{figure}

To illustrate the dyad vulnerability, we sampled 10,000 parameter vectors from $[-10,10]$ interval for both \texttt{edges} and \texttt{nsp(0)} at $v=7$. This region covers both the stable cone and the unstable region. We ran a 5000 step random walk MCMC trajectory for each sampled parameter until the first change occurred.  The simulation result (in fig.~\ref{fig:dyad_vul_simulation}) is in accordance with the theoretical derivation.

\section{Lazega’s Lawyer Dataset}
\label{sec:lazega}
In this section, we analyze network stability with a real-world dataset and a published model. We note at the outset that although perfect local stability (i.e., the stable cone w.r.t. $H^{(1)}$ is non-empty) is reached with the star example, not all model families lead to non-empty regions of local stability for $S=H^{(1)}$. For instance, consider a network with an attribute that classifies nodes to be in either group $A$ or $B$, and a model family that has two terms, a general \texttt{edges} term to control the overall density and a \texttt{nodemix} term that counts the number of between-group ties. If there exist vertices $(a,a')\subseteq A$ and $(b,b')\subseteq B$ such that edge $\{a,b\}$ is in $G$ and edge $\{a',b'\}$ is not in $G$, then the change scores for toggling the $\{a,b\}$ and $\{a',b'\}$ dyads have respective elements  \texttt{edges} = -1, \texttt{nodemix(A,B)} = -1 and \texttt{edges} = 1, \texttt{nodemix(A,B)} = 1. These two change score vectors define two half-spaces whose intersection is empty (i.e., $x+y > 0$ and $-x-y > 0$), and thus there is no model that stabilizes such a graph.   Intuitively, this is because any choice of $\mtheta$ for this family will on balance either favor adding or removing cross-group edges, and hence one of the two dyad states will not be favored.  More generally, for any $G$ and model family parameterized $\mt$, if there exist an edge and a null in $G$ with directly opposite change scores with respect to $\mt$, then the local stable region associated with $S=H^{(1)}$ will be empty.  In such cases, we can conclude that the forces governing the associated with $\mt$ do not (or would not) locally stabilize $G$.  Beyond this observation, we can gain additional insight into model behavior by examining the subsets of $S$ for which stabilization \emph{is} possible, particularly where $\mtheta$ is known or has been estimated from the observed network \cite[see e.g.][]{hunter.et.al:jcgs:2012}.  Moreover, where the local stable cone is non-empty but an estimated model does not lie within it, we can exploit the position of the estimated model relative to the stable cone to obtain insights into the factors that are driving instability, and the hypothetical changes in social forces that would lead the target graph to become locally stable.  In this section, we illustrate how some of these techniques can be used to gain insights into model behavior in a non-trivial setting.

We study graph stability with a data set collected by Lazega \cite{lazega2001collegial} on working relations among 36 partners in 1991 in a New England corporate law firm. This dataset is a network where edges (undirected) represent collaborations between partners. For purposes of analysis, we employ a model family for this data set that was previously published by \cite{hunter2007curved}. Due to improvements in estimation methods since the original publication, we here refit the model using the \texttt{ergm} package \cite{ergm3.8.0} to obtain updated coefficients (table \ref{tab:lazega_model}). The covariates used in the model are as follows: seniority, which describes the rank order of entry into the firm (1=earliest, 36=latest); type of practice (1=litigation, 2=corporate); the office at which the partner works (1=Boston; 2=Hartford; 3=Providence); and the partner's gender (1=man; 2=woman).  The model includes the main effects of seniority and practice, along with homophily effects for practice, sex, and office location.  A geometrically weighted edgewise shared partner (GWESP) term was also included to account for triadic closure.

\begin{table}[ht]
\centering
\begin{tabular}{lcl}
  \hline
 Parameter & Estimate & S.E \\ 
  \hline
Edges & -7.375 &  0.712$^{***}$ \\ 
  Main Seniority & 0.024 & 0.007$^{***}$ \\ 
  Main Practice & 0.411 & 0.118$^{***}$ \\ 
  Homophily Practice & 0.761 & 0.192$^{***}$ \\ 
  Homophily Gender & 0.696 & 0.256$^{**}$ \\ 
  Homophily Office & 1.145 & 0.196$^{***}$ \\ 
  GWESP($\alpha = 0.75$) & 0.937 & 0.159$^{***}$ \\   \hline
  \multicolumn{3}{c}{${}^{*}~p<0.05,~~{}^{**}~p<0.01,~~{}^{***}~p<0.001$}\\
\hline
\end{tabular}
\caption{Maximum likelihood estimates for the Lazega model, with standard errors and Wald test $p$-values.  $\alpha$ refers to the GWESP decay parameter, which was fixed to 0.75 during estimation.}
\label{tab:lazega_model}
\end{table}

\begin{figure}[h!]
\centering
  \includegraphics[width=6.5cm]{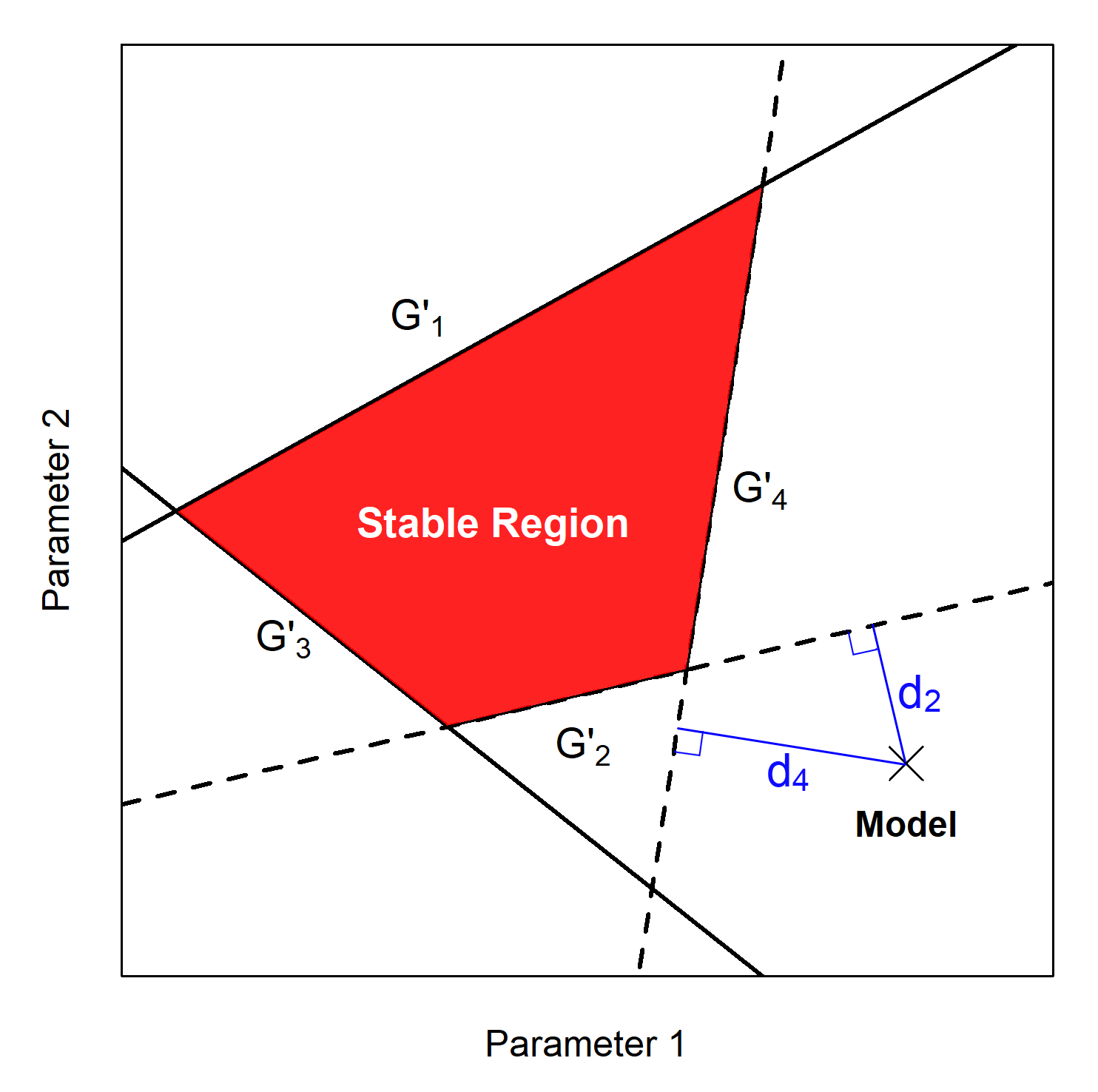}
  \caption{Example parameter space illustrating the stable region (in red), a particular model (black $\times$), and its distances to nearby hyperplanes (blue lines). Here, the target graph $G$ is unstable under the model, as the model is positioned outside of the stable region. Nevertheless, the model is partially stable w.r.t. $G'_1$ and $G'_3$, because it lies on the side of the stable half-spaces of these two alternative graphs (bottom of $G'_1$ and left of $G'_2$). In order to quantify the instability of the model, one can calculate the generalized distance from the model's position in the parameter space to each of the hyperplanes bounding the halfspaces for which the model is not included. }
  \label{fig:dist-illustration}
\end{figure}

We define $S=H^{(1)}$ (total 630 graphs), and compute the corresponding $M$ matrix (630-by-7). Each row in $M$ is characterized by one graph in the alternative set, and each column is one statistic. The network is locally stable under the estimated model with parameter $\hat{\mtheta}$ if $\mM\hat{\mtheta}  \in \mathbb{R_-}$, i.e. if all 1-step changes are unfavorable. Interestingly, the observed graph is unstable against this alternative set, with approximately and the unstable fraction is 0.159 (100 out of 630 graphs). Because each graph in the alternative set corresponds to one dyad toggle and defines a stable half-space, then for each dyad there exists a stable half-space in which this dyad is more likely to stay unchanged. If a model lies within the stable half-space, then the dyad is said to be stable under this model, and vice versa. We calculate the distance from the model to stabilization of graph $G'_i$ in $H^{(1)}$ as $d_i$ (shown in fig.~\ref{fig:dist-illustration}) and use the convention that positive $d_i$ means the model lies outside of the stable half-space of graph $G'_i$; while negative $d_i$ means the model is within the stable half-space of graph $G'_i$. The absolute value of $d_i$ is the distance to the dividing hyperplane. 

\begin{figure}[h!]
\centering
  \includegraphics[width=15cm]{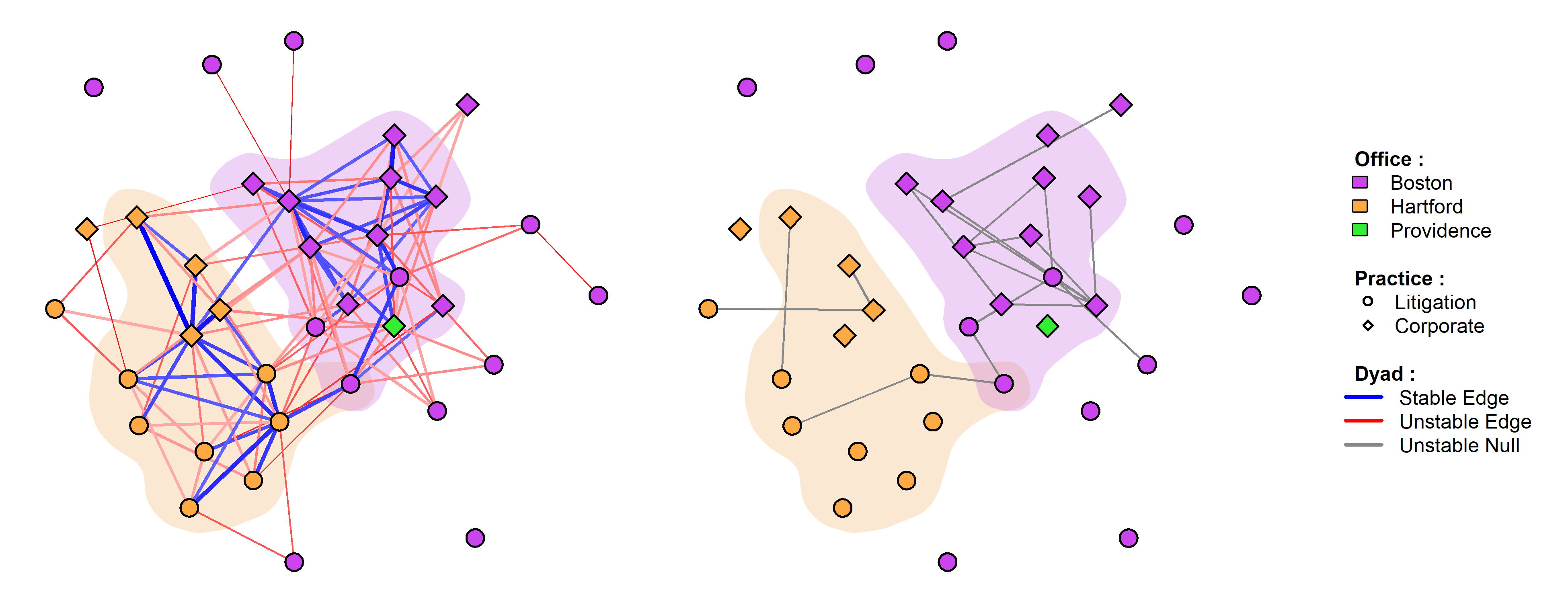}
  \caption{Illustration of the Lageza lawyer network with stable/unstable edges and nulls under the model. Specifically, edge stability is illustrated on the left panel, where the usage of color indicates stability. Unstable nulls are plotted on the right panel, with other nulls are stable.}
  \label{fig:lazega-stable-edges}
\end{figure}

In fig.~\ref{fig:lazega-stable-edges} we plot the stable edges, unstable edges, and unstable nulls. A few observations can be made: 1) Two core groups are clearly identified, each centered within one of the larger offices (Hartford and Boston). Ties within groups are generally more stable (in blue) and ties between groups are mostly unstable (in red). 2) Both large offices demonstrate core-periphery structures internally. The cores (marked by shaded background) are characterized with stable connections, while the peripheries are connected by unstable connections. 3) There are relatively few unstable nulls, and most of them are within two dense groups. In fact, there is only one unstable null that connects two groups. 4) Within two offices, there are a decent amount of unstable nulls in the Boston office and only a few in the Hartford office, suggesting that the pressure for forming new connections is less in the Hartford office. 

This analysis provides a basis for predicting what changes in the network can be expected. The external unstable edges between two groups suggest that cross-office ties tend to be somewhat fragile, and prone to disruption. The internal unstable edges connecting peripheral partners to core partners within groups likewise suggests an enhanced propensity towards turnover for collaborations involving these more marginal partners. By contrast, the unstable \emph{nulls} among core actors suggest the potential for new collaborations among the most central partners, especially in the Boston office. 
 
\begin{figure}[h!]
\centering
  \includegraphics[width=10cm]{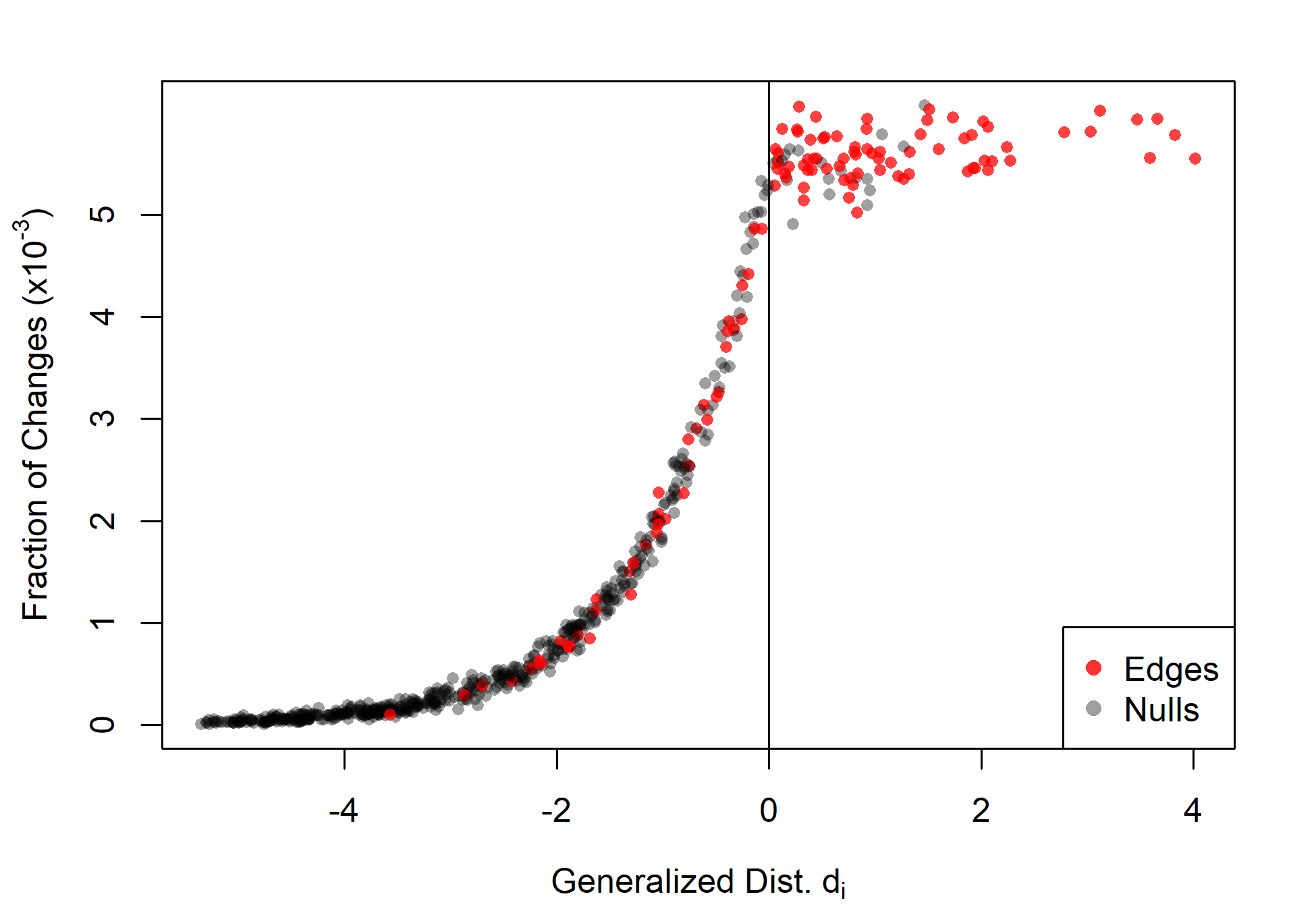}
  \caption{Dyad toggle occurrence fraction as a function of generalized distance $d_i$. Points on the left panel ($d_i < 0$) indicates dyads within the stable region, and points on the right ($d_i>0$) indicate unstable dyads. Edges and nulls are differentiated by color. }
  \label{fig:lazega-dist}
\end{figure}

The local stability calculations identify the dyads that are most vulnerable to changes in the network, i.e., if changes \emph{were} to happen, the unstable edges are the ones that would be expected to change the earliest. To see how this corresponds with an explicit dynamic process, we run 100,000 Metropolis trajectories until the first dyad toggle is accepted and record the number of times each dyad was toggled. We plot the fraction of each dyad toggle occurrence as a function of $d_i$ for all dyad $D_i$.  When $d_i < 0$ (left panel of fig.\ref{fig:lazega-dist}), indicating dyad $D_i$ is stable and the probability of accepting an edge toggle is proportional to the ratio of target/alternative graph potentials. When $d_i > 0$ (right panel of fig.\ref{fig:lazega-dist}), dyad $D_i$ is unstable and the fraction of dyad toggles becomes flat and is equal to the probability of any dyad being sampled ($\tbinom{v}{2}^{-1}$, which is 1/630 in this case). This reflects the fact that the Metropolis acceptance probability becomes 1 when a move is favored. This experiment shows that the most vulnerable dyads are the ones that are outside of the stable region, where the probability of such a dyad toggle is equal to the probability of any dyad being sampled. When a dyad lies within the stable region, the probability of a toggle is an exponentially increasing function of negated distance to the closest hyperplane, maxed out at dyad sampling probability. This indicates the dyads lie close to the hyperplanes, although within the stable region, are also somewhat vulnerable to structural changes. It is also interesting to note that under this model there are more unstable edges than nulls, indicate a tendency towards lowering network density by breaking established edges. 

\section{Discussion}

In this paper, we proposed a novel approach to analyze graph stability, based on the relative favorability of a target graph vis-\`a-vis a set of alternative graphs under a probability model. The requirements for applicability of this approach are very general, being merely a target graph, a set of alternative graphs against which stability is to be assessed, and a model class that parameterizes network probability as a monotonically increasing function of the linear combination of parameter values and graph statistics ($\mtheta^T \mt(g)$). ERGMs are a widely used model class of such kind, with the probability of a graph being proportional to the exponentiated linear terms. The stable region, if it exists, is the region where the target graph is more probable than any of the proposed alternative graphs. We construct an $|S|\mbox{-by-}K$ matrix $M$, each row of which is the generalized change score between the target graph and one of the alternative graphs (defined as $t(G')-t(G)$). Under this setting the stable set of a model is $\theta: \mM\theta \in \mathbb{R}^{|S|}_-$. We show that the stable region (if exists) is the interior of a convex $K$-polytope cone that points at the origin, and each facet of the cone is a row in $M$. We show an intuitive and easily implemented method to convert the facet representation to the ridge representation, whose worst-case complexity is the same as the exhaustively searching through all possible intersections. Nevertheless, the best case or the average case complexity is a massive speedup over the exhaustive method. 

Although this definition applies broadly to any target graph and arbitrary alternative set, certain alternative sets are particularly useful for providing intuition regarding model behavior and potential dynamics. For example, if we construct the alternative set to be all graphs that are Hamming distance one from the target graph, then stability against this set indicates that any one-step change is disfavored. For a random walk algorithm that assumes changes happen in series (i.e., one change at a time), this very local form of stability is an easily calculated approximation for dynamic stability (since any longer trajectory must still begin with a single step). 

To demonstrate our approach we showed a simple social example - an isolated star graph - inspired by the structures arising in certain charismatic cult groups.  Local stability ($S=H^{(1)}$) is particularly immediate in this case, as $S$ includes only two isomorphism classes.  Under a minimal model family, we verify that the stable region corresponds closely to the region that remains dynamically stable under random walk MCMC trajectories, with dynamic stability fading as one approaches the facets of the stable region. We further investigate the potential differences of graphs in $S$ near the facet of the stable region and show how dynamic stability is related to the potential difference of the alternative graphs versus the target. When $S$ is chosen to be $H^{(1)}$, any instability w.r.t. $S$ can also be interpreted in terms of dyad vulnerability, that is, under the current model forces, if changes were to happen to the network, which dyad is most vulnerable to being toggled. We divide the parameter space into regions that represent the types of dyads that are most likely to undergo changes from both theoretical derivation and simulation. 

We demonstrate a straightforward approach to exploring network stability under a given model, using the Lazega Lawyer dataset as an example. By simply constructing the $\mM$ matrix and examining whether the vector $\mM\hat{\mtheta}$ is in the negative quadrant, i.e., $\mM\hat{\mtheta} \in \mathbb{R}^{|S|}_-$, we can assess whether the observed graph is predicted to be stable under the estimated model. When the graph is not locally stable, then we may further inquire into the stability of particular dyads (exploiting the relationship between dyad toggles and the elements of $S$). Thus, non-negative elements in $\mM\hat{\mtheta}$ suggest the instability of the corresponding dyads. This can aid in making predictions regarding potential future changes to the network under the current social forces (as parameterized by the estimated model). For the Lazega dataset, we are able to identify the structural characteristics of the network from dyad stability assessment. For example, by examining the stable edges we are able to identify two densely connected clusters centered at two of the larger offices; each exhibits a core-periphery structure. The unstable nulls are relatively concentrated in the Boston office, suggesting that the pressure to collaborate is greater in the Boston office, while unstable edges bridging offices suggest fragility in ties between units. Such insights may be useful for guiding additional empirical studies or modeling efforts.

The methodology introduced here offers a powerful new toolset for practitioners of network modeling. The techniques presented require no simulation, and are applicable to a wide range of problems. The one-step stability metric introduced herein is a straightforward implementation of the alternative set, and possess an intuitive interpretation.  At the same time, our method is amenable to any user-defined alternative set. Our approach also offers a quantitative tool for measuring instability in network structure due to either a strain imposed on a network by the social forces at play, or conversely, could indicate that the choice of social forces in the model could be poorly chosen. Finally, we also note that the formal correspondence between the ERGM form and Boltzmann distribution makes this approach useful in physical settings \cite{grazioli.et.al:jpcB:2019}, where locally stable structures correspond to local energy minima in graph space.  The ability to easily characterize the conditions under which particular graph structures are energetically favorable may be useful for studying the formation of complex materials, or the protein aggregates associated with Alzheimer's and other diseases.


\end{document}